\documentclass[leqno,11pt]{article}

\usepackage{graphicx, wrapfig, subfig} 
\usepackage{amsmath, amsfonts} 
\usepackage{amsthm} 
\usepackage{amssymb} 
\usepackage{mathrsfs} 
\usepackage[retainorgcmds]{IEEEtrantools} 
\usepackage{xcolor} 

\graphicspath{{./Figures/}{./matlab/}{./}} 

\newcommand{\abs}[1]{\left\lvert #1 \right\rvert}

\DeclareMathOperator{\sign}{sign}

\theoremstyle{plain}

\theoremstyle{definition}

\theoremstyle{remark}

\newcommand{\Sfrac}[2]{{\textstyle \frac{#1}{#2}}}

\title{Numerical Bifurcation for the Capillary Whitham Equation}
\author{Filippo Remonato\footnote{Department of Mathematical Sciences, NTNU, 7491 Trondheim, Norway,
filippo.remonato@math.ntnu.no}~
 and Henrik Kalisch\footnote{Department of Mathematics, University of Bergen,
        5020 Bergen, Norway, henrik.kalisch@math.uib.no}}

\begin{document}

\maketitle

\begin{abstract}
The so-called Whitham equation arises in the modeling of free surface water waves, 
and combines a generic nonlinear quadratic term with the exact linear dispersion relation 
for gravity waves on the free surface of a fluid with finite depth. 

In this work, the effect of incorporating capillarity into the Whitham equation is in focus. 
The capillary Whitham equation is a nonlocal equation similar to the usual Whitham equation,
but containing an additional term with a coefficient depending on the
Bond number  $T$ which measures the
relative strength of capillary and gravity effects on the wave motion.

A spectral collocation scheme for computing approximations to periodic traveling waves 
for the capillary Whitham equation is put forward. Numerical approximations of periodic traveling waves 
are computed using a bifurcation approach, and a number of bifurcation curves are found. 
Our analysis uncovers a rich structure of bifurcation patterns, 
including subharmonic bifurcations, as well as connecting and crossing branches.
Indeed, for some values of the Bond number $T$, the bifurcation diagram features 
distinct branches of solutions which intersect at a secondary bifurcation point.
The same branches may also cross without connecting, and some bifurcation curves 
feature self-crossings without self-connections.
\end{abstract}

\section{Introduction}
\label{sec:1}
The Korteweg-de Vries (KdV) equation 
\begin{equation}\label{eq:kdv}
\eta_t + c_0 \, \eta_x + \frac{3}{2}\frac{c_0}{h_0} \, \eta \, \eta_x 
                    + \frac{1}{6} c_0 h_0^2 \, \eta_{xxx} = 0
\end{equation}
is a simplified model equation for waves at the surface of a fluid
contained in a rectangular channel.
The equation includes the competing effects of nonlinear steepening
and frequency dispersion \cite{KdV}.
Balancing these two effects is the basic mechanism behind the
existence of both solitary-wave solutions and periodic travelling
waves.
Equation \eqref{eq:kdv} is given in dimensional form,
$c_0 = \sqrt{g h_0}$ is the limiting long-wave speed,  
$h_0$ denotes the undisturbed water depth, 
and $g$ is the gravitational constant of acceleration. 
The function $\eta(x,t)$ describes the deflection 
of the fluid surface from the rest position at a point $x$ at time $t$.
The equation is a valid approximation describing the evolution
of surface water waves in the case when the waves are long
compared to the undisturbed depth $h_0$ of the fluid, the average amplitude
of the waves is small when compared to $h_0$ 
and in addition, transverse effects are assumed to be weak
\cite{BCL,C,LannesBOOK,Wh2}.

The linear phase speed in the KdV equation is given by
\begin{equation}\label{dispKdV}
c(\xi) = c_0 - \frac{1}{6} c_0 h_0^2 \xi^2,
\end{equation}
where $\xi=\frac{2 \pi}{\lambda}$ is the wave number, and $\lambda$
is the wavelength. 
This is a second-order approximation to the wave speed
\begin{equation}\label{disp}
c(\xi) = \frac{\omega}{\xi} 
       = \sqrt{{\textstyle \frac{g \tanh{\xi h_0}}{\xi} }},
\end{equation}
of the linearised water-wave problem. 
The latter expression for $c(\xi)$ appears when the full water-wave problem 
is linearised around the vanishing solution, and solutions of the form 
$\exp(ix \xi - i \omega t)$ are sought \cite{Wh2}. 

Comparing the expressions \eqref{dispKdV} and \eqref{disp}, it appears that the linearised KdV equation does not
give a faithful representation of the full dispersion relation even for intermediate values of the wave number $\xi$.
Recognising this problem of the KdV equation as a model equation for water waves, Whitham introduced
what is now called the Whitham equation \cite{Wh1}.
The idea was to use the exact form of the wave speed \eqref{disp}
instead of a second-order approximation like \eqref{dispKdV}.
The equation proposed by Whitham has the form
\begin{equation}\label{eq:whit}
\eta_{t} + \frac{3}{2}\frac{c_0}{h_0} \, \eta \, \eta_{x} 
         + K_{h_0} * \eta_{x} = 0,
\end{equation}
where the convolution is in the $x$-variable.
The equation is written in dimensional variables,
with $\eta(x,t)$ representing the deflection of the surface from rest, 
just as in the KdV equation.
The convolution kernel is defined via the Fourier transform $\mathcal{F}$ by 
\begin{equation}\label{eq:Kh}
\mathcal{F} K_{h_0} = c(\xi) = \sqrt{\textstyle{ \frac{g\tanh h_0 \xi}{\xi}}}.
\end{equation}
It should be mentioned that the Whitham equation has excited some interest
because it was conjectured to feature wave breaking and peaking.
Wave breaking in this context is defined as the development of an 
infinite gradient in the solution. In a physical context, this kind
of breaking may not happen naturally for a free equation such as \eqref{eq:whit},
but may require some forcing either by a sloping bottom, or an imposed
discharge \cite{BjKa2011}. 
While the KdV equation does not allow the formation of infinite gradients,
it features convective wave breaking which is related to spilling at the wavecrest \cite{BrKa2016}. 
Wave peaking describes the situation where
a steady wave profile features a singular point, such as a peak or 
a cusp, such as in the well known highest wave which was conjectured
to be peaked by Stokes, and proved to exist in \cite{AFT82,Plotnikov}.

Both the existence of peaked and breaking waves were investigated
to some degree already by Whitham \cite{Wh1,Wh2}, and studied at
length for a number of related equations by Naumkin and Shishmarev in the monograph \cite{NS}.
Recently, proofs of both phenomena have become available. In particular, it was shown
in \cite{Hur2015} that the Whitham equation features waves which develop an infinite
gradient, and the existence of a highest, peaked wave was proved in \cite{EW2016}.

In the present article, the Whitham equation is studied in the case when surface tension is important.
The motivation for this pursuit lies partially in the analysis in \cite{MKD2015} where
it was shown that the Whitham equation is a valid model for surface waves of smaller wavelengths
than the KdV equation. As a result, it is possible to use the Whitham equation for surface waves
which are short enough for capillary effects to play a role. On the other hand, there are situations
where capillarity is strong, such as in the presence of a surface film
or an interfacial hydrate layer \cite{Be3,HH2005,Kalisch2007}.
In this case, capillarity can be important even for longer waves.

In the general case where both capillary and gravity effects are present, 
the relation between the wavenumber $\xi$ and the radial frequency $\omega$ 
in the linearised surface water wave problem is given by
\begin{equation}
\omega^2 = g \xi \tanh(\xi h_0) \left( 1 + \Sfrac{\tau}{\rho g} \xi^2 \right),
\end{equation}
where $\rho$ is the density of the fluid, and $\tau$ is the surface tension
of the free surface.

If restricted to waves propagating into a single direction, the phase velocity
can be written as
$$
c(\xi) = {\textstyle \sqrt{ \frac{g\tanh h_0 \xi}{\xi} \big( 1 + \frac{\tau}{\rho g} \xi^2 \big)}}.
$$
Thus in the case of capillary-gravity waves, this definition of $c(\xi)$ is used
in the definition of the integral kernel in \eqref{eq:Kh}.
If the undisturbed depth $h_0$ is taken as a unit of length, and $h_0 / c_0$ is taken as unit of time,
then the Whitham equation with surface tension is
\begin{equation}\label{eq:whitT}
u_t + \frac{3}{2} u u_x + K_T * u_x = 0,
\end{equation}
where the integral kernel $K_T$ is given by its Fourier transform, viz.
\begin{equation}\label{K_T}
\mathcal{F} K_T(x) = {\textstyle \sqrt{\frac{(1+T\xi^2)\tanh(\xi)}{\xi}}},
\end{equation}
where $T=\frac{\tau}{\rho g h_0^2}$ is the Bond number which measures the
relative strength of capillary and gravity effects on the wave motion.
In particular, for $T=0$ one recovers the purely gravitational case.

Note that this equation is completely different in structure from the
capillary KdV equation
\begin{equation}
u_t + c_0 \, u_x + \frac{3}{2} u u_x 
                    + \frac{1}{6} \, u_{xxx} - \frac{T}{2} u_{xxx} = 0.
\end{equation} 
This latter equation reduces to the case of the KdV equation with the sign of the dispersive term being positive
or negative depending on the value of $T$. Since these two cases are equivalent via a change of sign, 
they do not differ in a qualitative way \cite{Be2}. 
The one case of greater interest is when the Bond number $T$ is close to $1/3$ 
as a fifth-order term is then needed in order to get the correct order of approximation. 
The resulting equation is known as the Kawahara equation, and it features competing third
and fifth order derivatives. On the other hand, equation \eqref{eq:whitT}
features two competing nonlocal terms for any value of the Bond number $T$,
and as will be seen presently, this configuration has repercussions on the
possible solutions of the equation.

In the present work, steady solutions of \eqref{eq:whitT} are under consideration and we will 
look for solutions in the space of continuous $2\pi/k$-periodic functions,
which will be denoted by $C_{2\pi/k}$.
For convenience, we use a further rescaling to put \eqref{eq:whitT} in the tidy form
\begin{equation}
u_t +  2 u \, u_x + K_T * u_x = 0,
\end{equation}
and then use the assumption \( \eta(x,t) = u(x-\mu t)\) to search for travelling wave solutions
with propagation speed $\mu$.
The equation can then be written
in integrated form as
\begin{equation}
    \label{eqWhitham}
    W(\mu, u) = -\mu\,u + u^2 + K_T*u = 0.
\end{equation}

As will be shown in the body of this article, with the definition of $K_T$ in \eqref{K_T},
equation \eqref{eqWhitham} features a large variety of solutions.
In particular, there are branches which contain secondary bifurcation points
leading to connections with other branches. There are also crossings
of distinct branches without connections, and
there are self-crossing (but not intersecting)
bifurcation branches. Such patterns have been seen before in some cases, such
as in the case of tri-modal surface water waves (\cite{EW2015}), but the nature
of the connections appears to be different in the present case. The existence
of crossing and self-crossing branches leads to non-uniqueness of solutions
of the steady problem \eqref{eqWhitham} which is an interesting problem in itself.

The plan of the paper is as follows. In Section \ref{secBifurcationFormulas},
analytic bifurcation formulae are provided in order to guide the numerical
experiments. In Section \ref{scheme}, the numerical scheme is explained in detail,
and in Section \ref{secResults}, numerical experiments are shown.

\section{Analytic expansions}
\label{secBifurcationFormulas}

We now want to provide an analytical expansion of the wave profile and speed near the bifurcation point. 
We look for an expansion in the form
\begin{align}
	\label{eqExpansionFormulas_u}
	u_\epsilon &= u_1\,\epsilon + u_2\,\epsilon^2 + u_3 \, \epsilon^3 + u_4 \, \epsilon^4 + \ldots \\
	\label{eqExpansionFormulas_c}
	\mu_\epsilon & = \mu_0 + \mu_1\,\epsilon + \mu_2\,\epsilon^2 + \mu_3 \, \epsilon^3 + \ldots \text{.}
\end{align}
In this pursuit, it is important to understand the behaviour of the dispersion relation 
in terms of different values of the Bond number $T$.

\subsection{Bifurcation speed}
Analysing the linearised version of \eqref{eqWhitham}, it is intuitively clear that
given $k\in\mathbb{N}$, the speed at which non-trivial $2\pi/k$-periodic solutions 
bifurcate from the trivial solution curve is given by
\begin{equation}
	\label{eqBifurcationSpeedCapillary}
	\mu^* = m(k) = \sqrt{\frac{(1+T k^2)\tanh(k)}{k}} 
\end{equation}
and the kernel of $D_u W$ at the bifurcation point is the span of $\{\cos(kx)\}$.
A firm proof of this fact can be established in the same way as it was shown
for the purely gravitational case in \cite{Ehrnstrom2009twf}.
\begin{figure}[t]
	\centering
	\subfloat[$T=0$]{
		\includegraphics[width=0.39\linewidth]{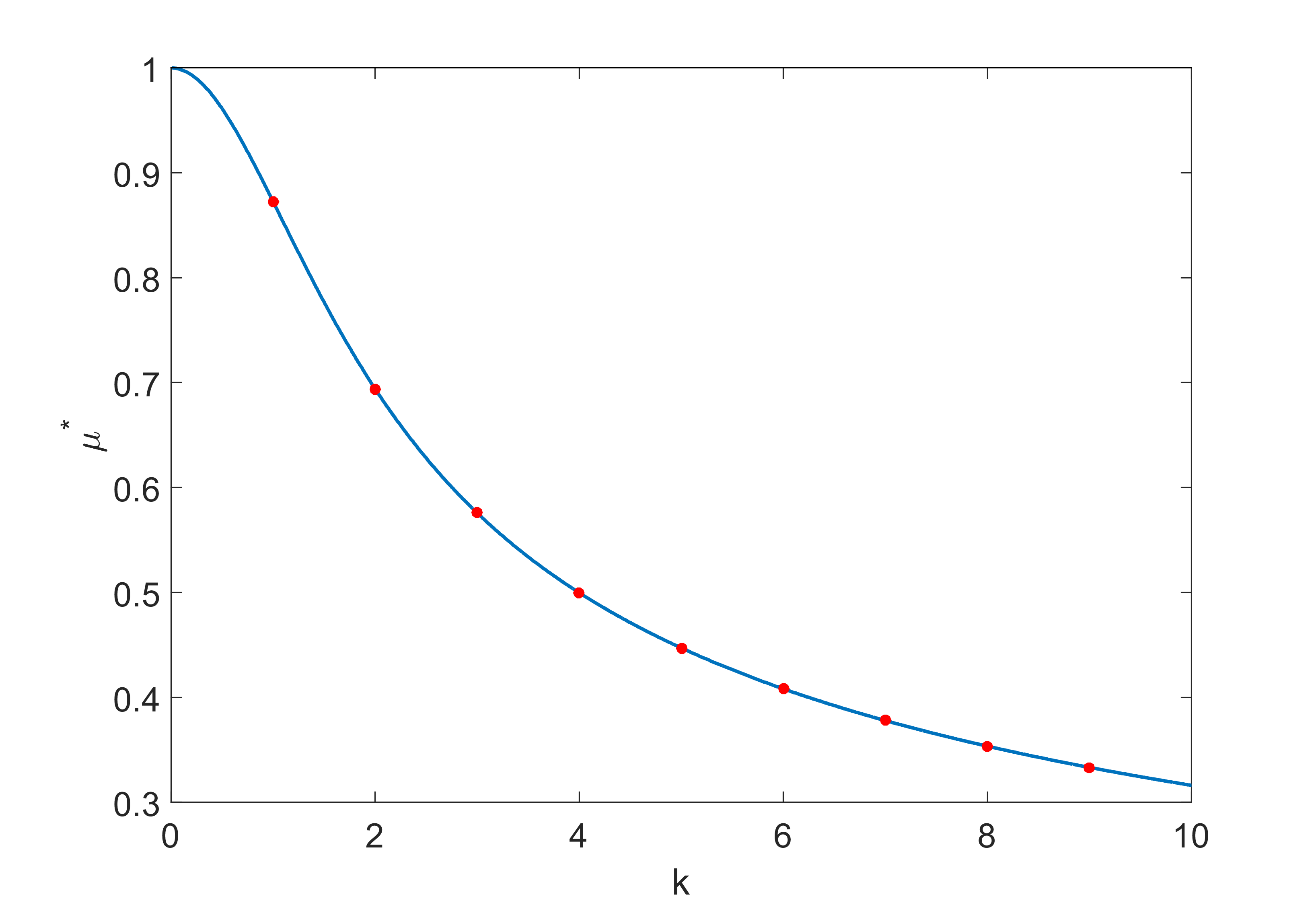}
	}
	\subfloat[$T=0.1$]{
		\includegraphics[width=0.39\linewidth]{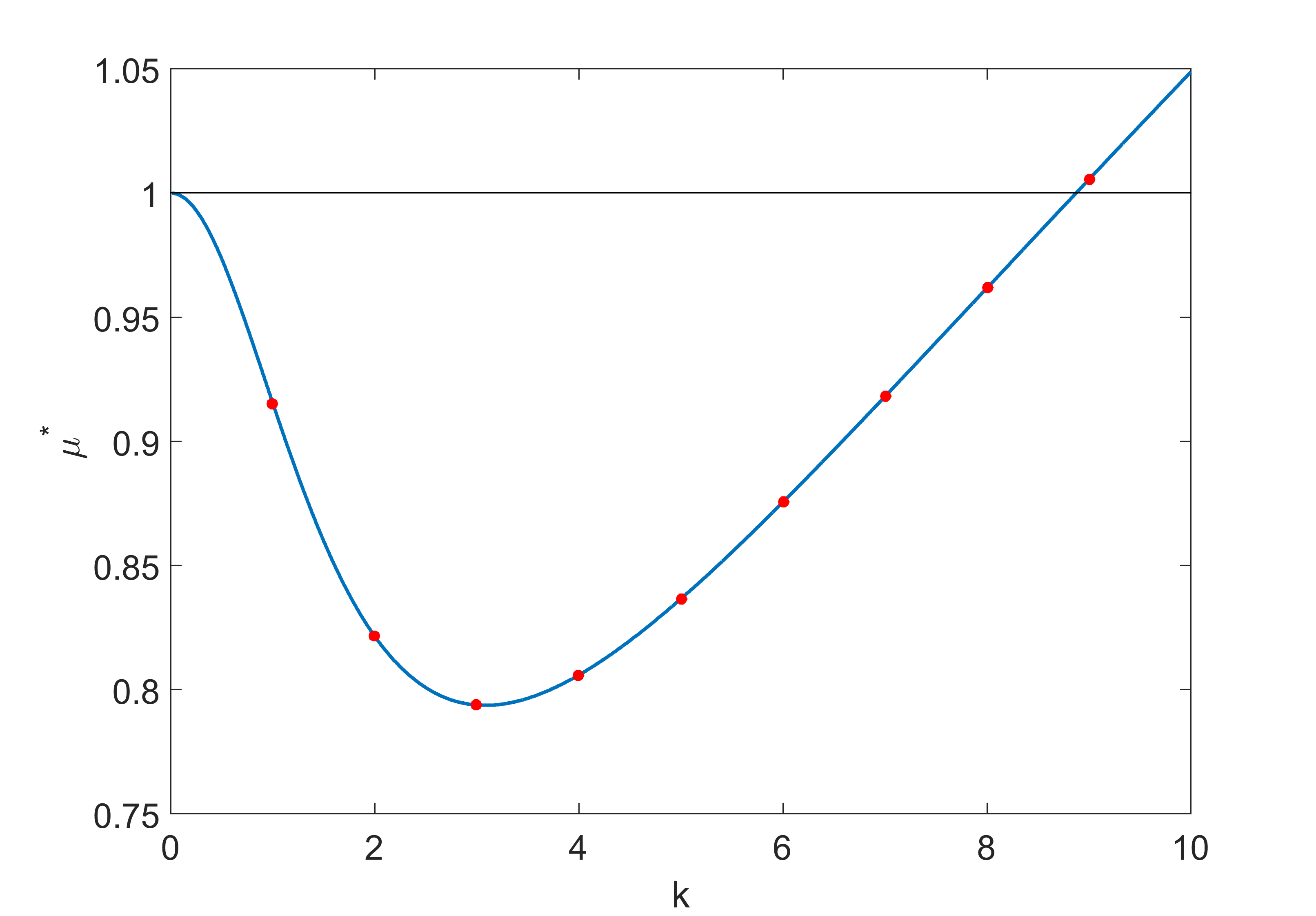}
	}
	
	\subfloat[$T=0.2$]{
		\includegraphics[width=0.39\linewidth]{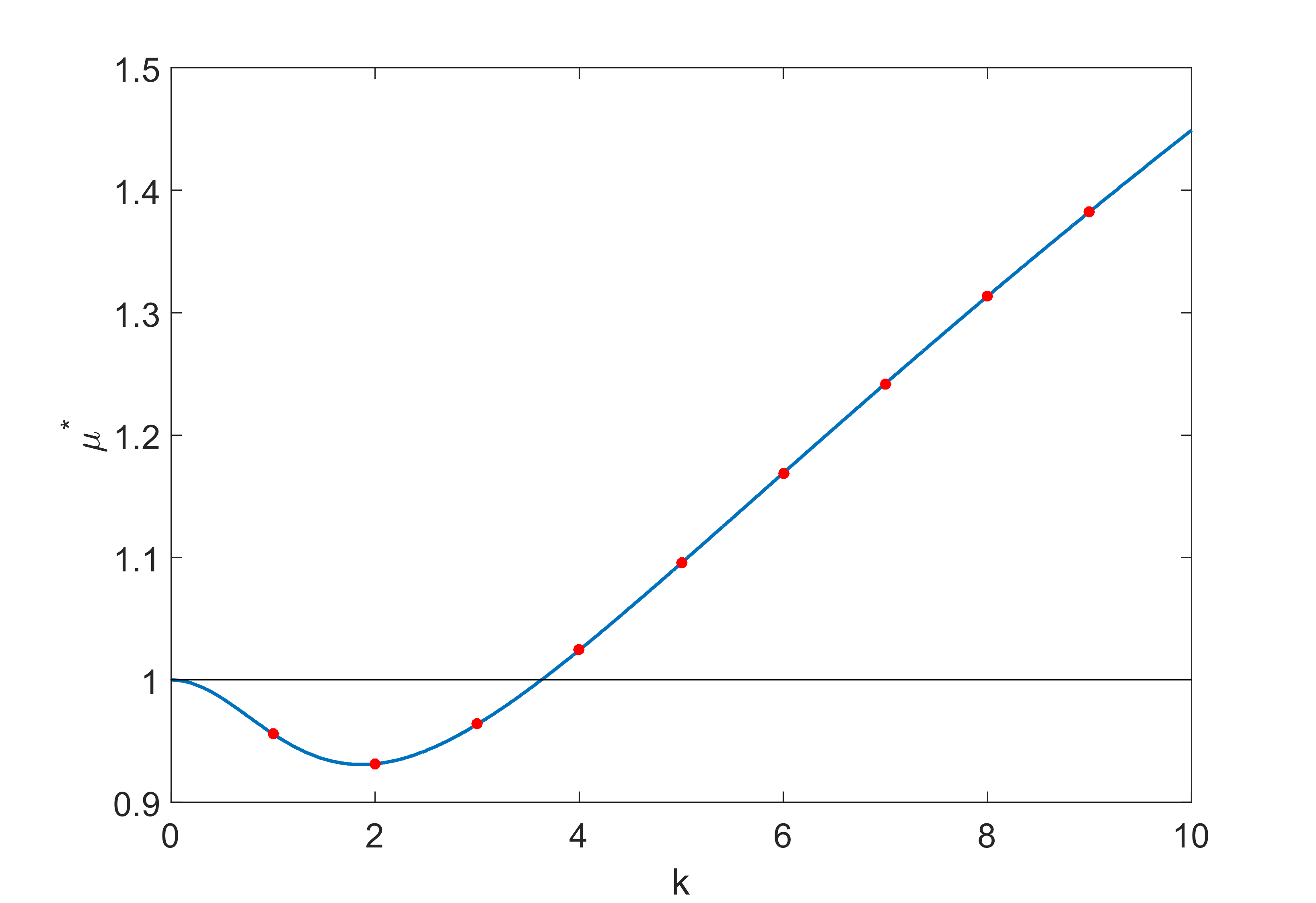}
	}
	\subfloat[$T=0.333$]{
		\includegraphics[width=0.39\linewidth]{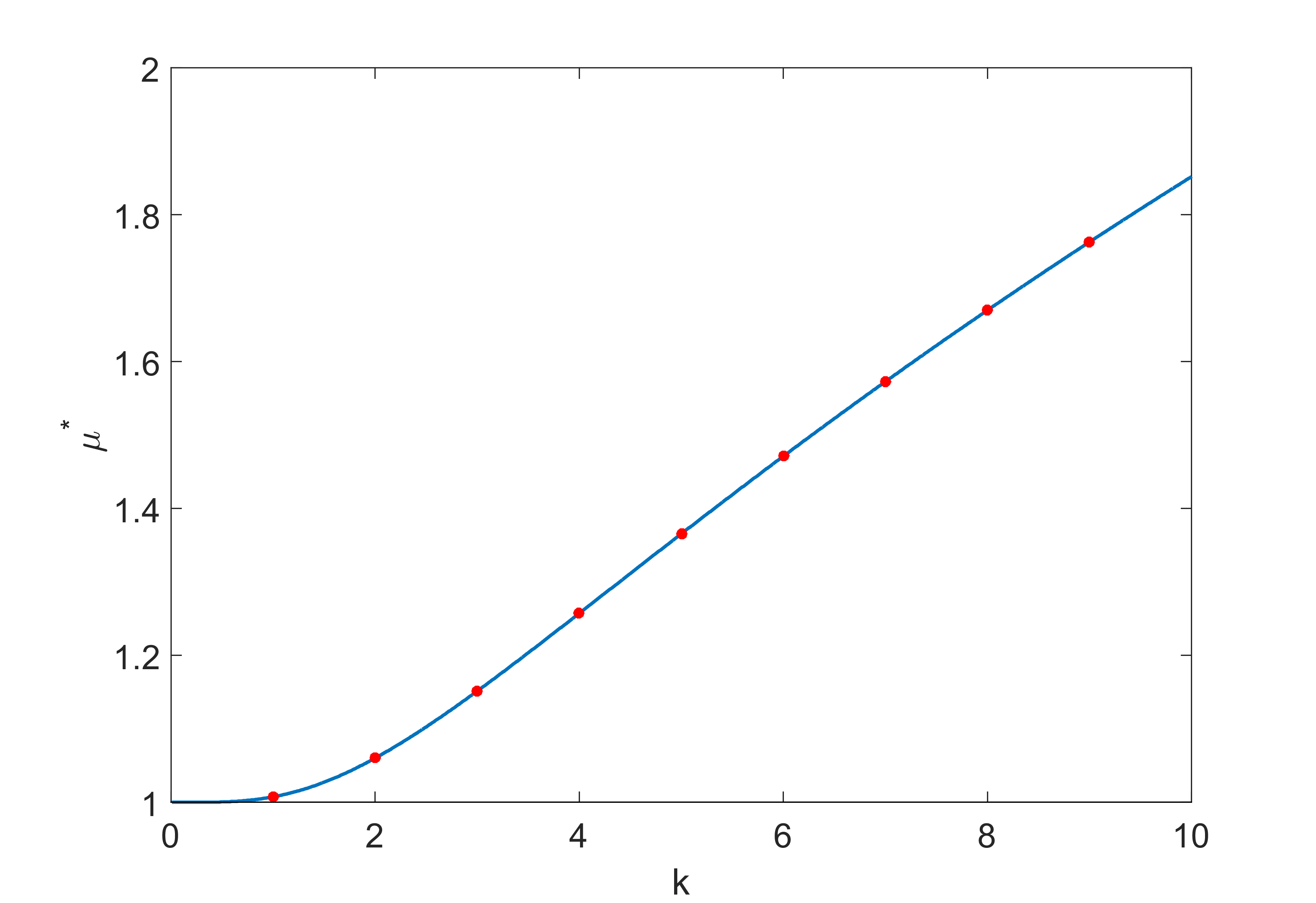}
	}
	\caption{\label{figBifurcationSpeed} 
The bifurcation speed $\mu^*$ as a function of the wave number $k$ for various values of the Bond number $T$. 
The case $T=0$ corresponds to the gravitational case of Equation \eqref{eqWhitham}. 
Panels (b) and (c) illustrate two cases where the dispersion curve is non-monotone. Panel (d) shows
the case where $T=1/3$. For $T\ge 1/3$, the curve is monotone.}
\end{figure}

It can be shown that for $T=0$, Equation \eqref{eqBifurcationSpeedCapillary} 
is monotonically decreasing in $k$,
while it has a global minimum for any value of $T>0$. In particular, $\min m(k) \in (0,1)$ for $0<T<\frac{1}{3}$,
while for $T\geqslant \frac{1}{3}$ the minimum is 1 and $m(k)$ is monotonically increasing in $k$.
Some examples are shown in {Figure \ref{figBifurcationSpeed}}.

This means that given two wavenumbers $k_1, k_2$, we can always find a $T$ such that $m(k_1) = m(k_2)$,
and hence the two branches bifurcate from the same point. Such $T$ is given by
\begin{equation}
	\label{eqTForCollision}
	T(k_1, k_2) = \frac{k_1 \tanh(k_2) - k_2 \tanh(k_1)}{k_1 k_2 \left(k_1\tanh(k_1) - k_2 \tanh(k_2)\right)}
\end{equation}

Note that this implies that for $T = T(k_1, k_2)$, the kernel of $D_u W$ is two-dimensional at the bifurcation point,
and in particular the kernel is the span of $\{\cos(k_1x), \cos(k_2x)\}$.
This fact, along with the existence of local \emph{sheets} of solutions, 
is outside of the scope of the present paper, but will
be rigorously proved in future work.

\subsection{Expansion coefficients and multi-modal waves}
\label{secExpansionCoeffs}
In the case of a one-dimensional kernel, i.e. when $T \ne T(k_1, k_2)$,
the constants in formulas \eqref{eqExpansionFormulas_u} and \eqref{eqExpansionFormulas_c} are given below:

\begin{IEEEeqnarray*}{rCl}
	u_1 &=& \cos(kx), \\
	u_2 &=& \frac{1}{2\,(m(k) - 1)} + \frac{1}{2\,(m(k) - m(2k))}\,\cos(2kx), \\
	u_3 &=& \frac{1}{2\,(m(k) - m(3k))\,(m(k) - m(2k))} \, \cos(3kx),\\[5pt]
	u_4 &=& A_0 + A_{2k} \cos(2kx) + A_{4k}\cos(4kx). \\
\end{IEEEeqnarray*}
The last function is defined in terms of the constants
\begin{IEEEeqnarray*}{rCl}
	A_0 &=& -\frac{1}{4\,(m(k) - 1)^3} - \frac{1}{8\,(m(k) - 1)^2 (m(k)-m(2k))} \\
		&& + \frac{1}{8\,(m(k) - 1)\,(m(k)-m(2k))^2}, \\
	A_{2k} &=& -\frac{1}{4\,(m(k)-m(2k))^3} + \frac{1}{4\,(m(k)-m(2k))^2 (m(k)-m(3k))}, \\[5pt]
	A_{4k} &=& \frac{1}{8\,(m(k) - m(2k))^2 (m(k)-m(4k))}, \\
		&& + \frac{1}{2\,(m(k) - m(2k))\,(m(k)-m(3k))\,(m(k)-m(4k))}. \\
\end{IEEEeqnarray*}
For the expansion of the wave speed $\mu$, we have
\begin{IEEEeqnarray*}{rCl}
	\mu_0 &=& m(k),\\
	\mu_1 &=& 0, \\
	\mu_2 &=& \frac{1}{m(k) - 1} + \frac{1}{2(m(k) - m(2k))},\\
	\mu_3 &=& 0.
\end{IEEEeqnarray*}

Note that these expansions coincide, up to the second order in $\epsilon$, with the bifurcation formulas given in \cite{Ehrnstrom2013gbf}.

Due to Formula \eqref{eqTForCollision} there exist some values of $T$ for which the above expansion is not valid, e.g. when $T=T(k,2k)$. In those cases a more in-depth analysis is required.
However, since all the terms in the denominator are of the form $(m(k)-m(a k)), a\in\mathbb{N}_0$, the expansions \eqref{eqExpansionFormulas_u} and \eqref{eqExpansionFormulas_c} remain valid also when $T = T(k_1,k_2)$ provided $k_2 \neq a k_1$.
In the other cases, we can still select $T$ in order for \eqref{eqExpansionFormulas_u} and \eqref{eqExpansionFormulas_c} to hold while making the coefficients in a component $u_n$ arbitrarily large.
This explains the existence of multi-modal waves, which are associated with the property that the bifurcation kernel can be two-dimensional. For instance, in \cite{EW2015}, tri-modal waves were found 
in the case of the full-water wave problem with a background shear current.
Several examples are presented in Section \ref{secResults}.

\subsection{Tangent and direction of nontrivial curves at the bifurcation point}
\label{secConcavity}
Due to Equation \eqref{eqBifurcationSpeedCapillary} it is natural to use the wave speed as a bifurcation parameter, and we are interested in the shape of curves of nontrivial
solutions close to the bifurcation point. This information is given by the expansion \eqref{eqExpansionFormulas_c}
for the wave speed except in the cases of a two-dimensional kernel.

In the purely gravitational case we know that any nontrivial branch has a vertical tangent at the bifurcation point.
This is due to the fact that $\mu_1=0$, and as we have have just shown it is preserved also in the capillary case.

\begin{figure}[htb]
	\centering
		\subfloat[$k=1$]{
		\includegraphics[width=0.39\linewidth]{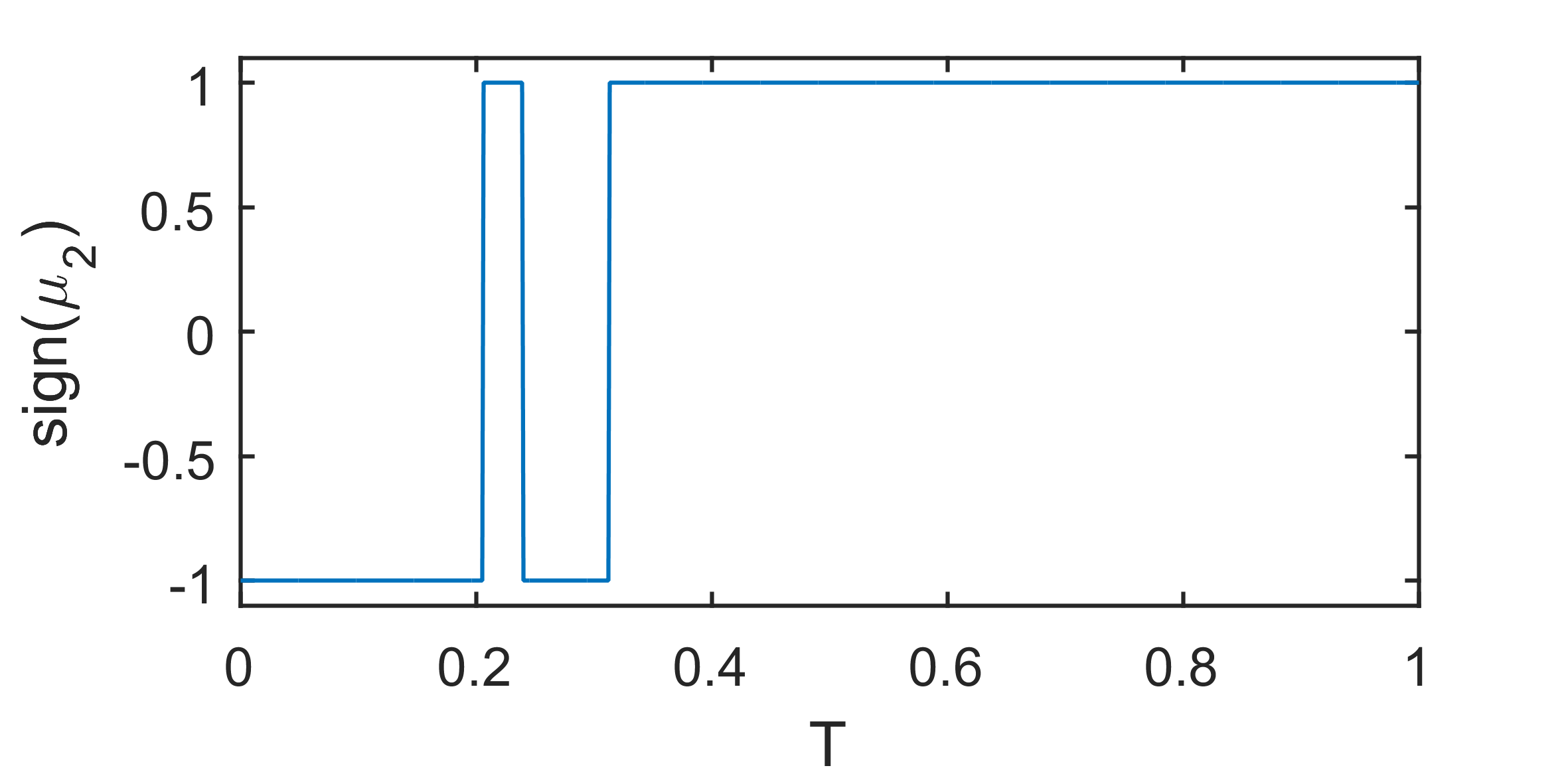}}
	\subfloat[$k=2$]{
          \includegraphics[width=0.39\linewidth]{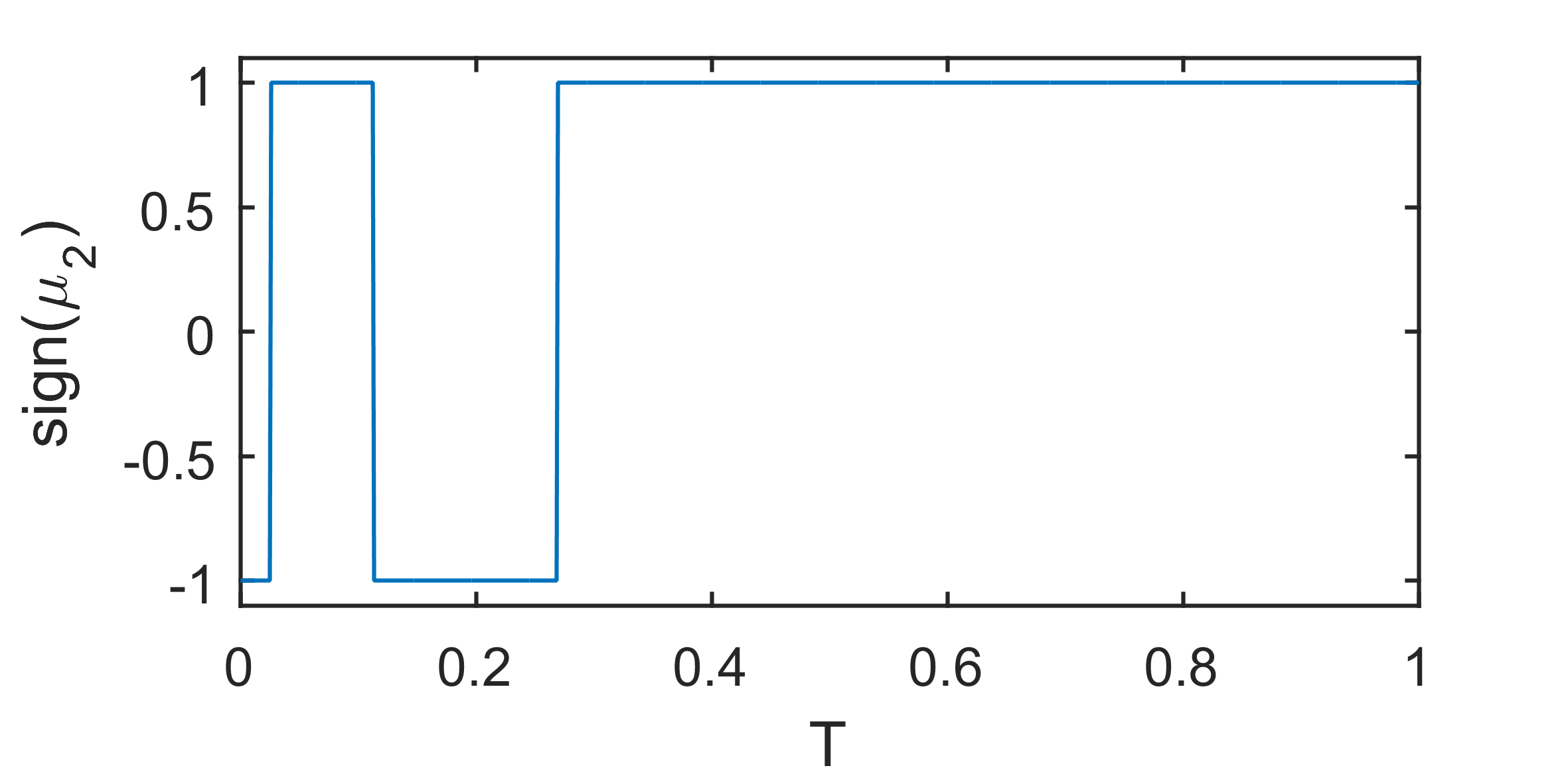}}
        
	\subfloat[$k=3$]{
		\includegraphics[width=0.39\linewidth]{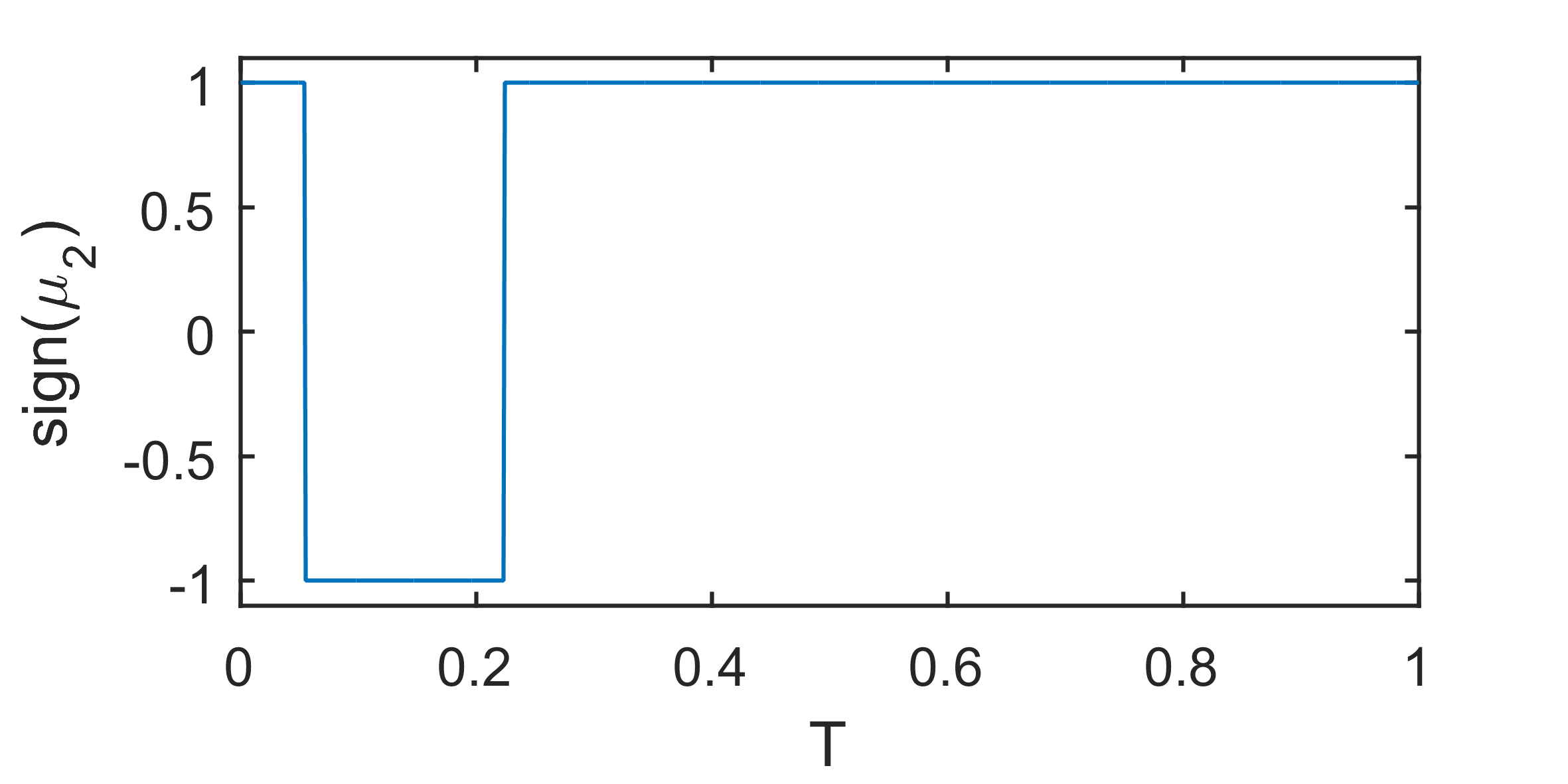}}
	\subfloat[$k=4$]{
		\includegraphics[width=0.39\linewidth]{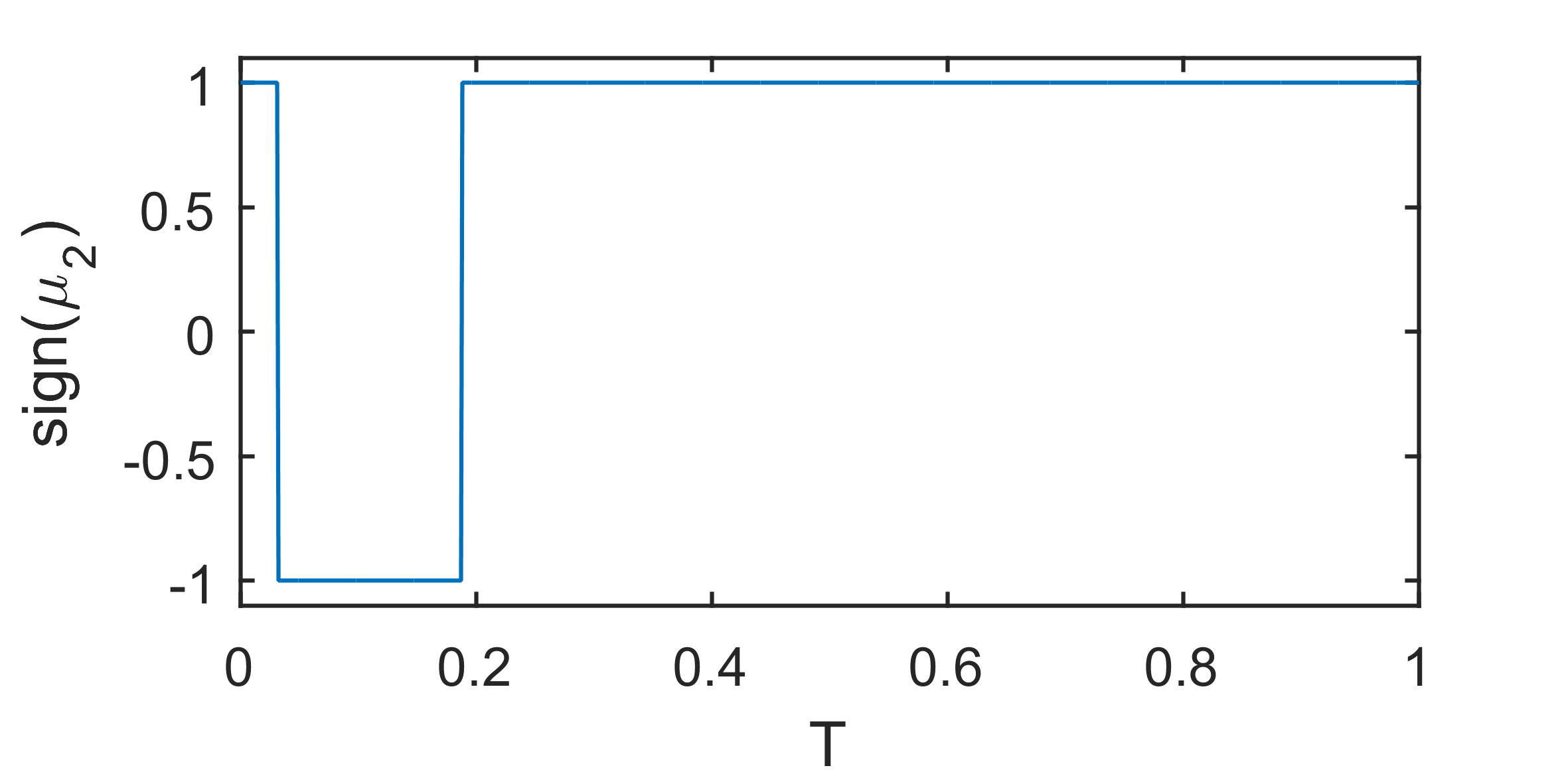}}
	\caption{\label{figConcavity}
          Values of $\text{sign}(\mu_2)$ for $T\in(0,1)$ and $k=1,\ldots,4$.
          Positive values mean the branch goes to the right of $\mu^*$, negative values that it goes to the left. }
\end{figure}

Moreover, for gravity waves it was shown in \cite[Theorem 4.6]{Ehrnstrom2013gbf}
that the main branch (k=1) satisfies $\ddot \mu(0) = \mu_2 < 0$,
which means that in a neighbourhood of the bifurcation point the main branch will go to the left,
in the direction of decreasing velocities.
Due to the effect of $T$ on $m(k)$, we can see from the above bifurcation formulas
that there are values of $T$ for which $\mu_2$ changes sign,
and therefore the main branch can bifurcate going to the right, in the direction of increasing velocities.
The value of $\text{sign}(\mu_2)$ is plotted in Figure \ref{figConcavity} for the first four wavenumbers.
Note that the branches for $k=3,4$, among others, bifurcate going to the right also in the purely gravitational case.

\section{The numerical scheme}
\label{scheme}

We employ a variation of the method presented in \cite{Ehrnstrom2013gbf}.
We want to apply a Fourier-collocation method, which is convenient given the definition of $K$.
Also note that, thanks to symmetry, we can perform all computations on the half-wavelength $L = \lambda/2 = \pi/k$.  
Given $k$, let $N$ be the total number of collocation points and define the subspace of $L^2(0,\pi)$
\begin{equation*}
	\mathcal{S}_h = \text{span} \{ \cos(nx) : 0 \leqslant n \leqslant N-1 \}
\end{equation*}
and the collocation points $x_i = \frac{(2i-1)\,\pi}{2Nk}$ for $i=1,\ldots,N$.
We then discretise Equation \eqref{eqWhitham} and search for a solution $u_h \in \mathcal{S}_h$, $u_h(x_i) = u_i$ such that
\begin{equation}
	\label{eqDiscreteEquation}
	-\mu \,u_h + u_h^2 + K \, u_h = 0
\end{equation}
To understand the term $K_h\,u_h$,
we need to see how $K$ acts on functions in $\mathcal{S}_h$, therefore we expand $u_h$ in its discrete Fourier (cosine) series:
\begin{equation}
    \label{eqCosineSeriesU}
    u_h(x) = \sum_{n=0}^{N-1} w_n\,a_n\cos(nx),\quad a_n = w_n\,\sum_{i=1}^N u_i\cos(kx_i),
\end{equation}
where as usual
\begin{equation*}
    w_n =
    \begin{cases}
        1/\sqrt{N} & n=0\\
        2/\sqrt{N} & n\geqslant 1.
    \end{cases}
\end{equation*}

We then see that $K$ acts on $u_h$ as follows:
\begin{align*}
  K*u_h &= \int K(y)\,u_h(x-y)\;\mathrm{d}y = \int K(y)\,\sum_{n=0}^{N-1} w_n\,a_n\,\cos(nx-ny)\;\mathrm{d}y \\[5pt]
    &= \int K(y) \sum_{n=0}^{N-1}w_n\,a_n\; \Sfrac{e^{i(nx-ny)}+ e^{-i(nx-ny)}}{2} \;\mathrm{d}y.
\end{align*}
We now split the integral in the two parts, change variables $y\mapsto-y$ in the second integral, 
and exploit the fact that $K$ is even, and get that the above becomes
\begin{align*}
K*u_h &= \sum_{n=0}^{N-1} w_n\,a_n\,\Sfrac{e^{inx}-e^{-inx}}{2}\;\int K(y) e^{-iny} \;\mathrm{d}y \\[5pt]
    &= \sum_{n=0}^{N-1} w_n\,a_n\,\cos(nx)\;{\textstyle \sqrt{ \frac{(1+Tn^2)\tanh(n)}{n}}}.
\intertext{Expanding the definition of $a_n$ and rearranging the sums we finally have}\\[-12pt]
    &= \sum_{i=1}^N \sum_{n=0}^{N-1} w_n^2 \;{\textstyle \sqrt{\frac{(1+Tn^2)\tanh(n)}{n}}}\;\cos(nx_i)\cos(nx) \; u_i.
\end{align*}
So if we define the matrix $\boldsymbol K$ as
\begin{equation*}
    \boldsymbol K(i,j) = \sum_{n=0}^{N-1} w_n^2 \;
{\textstyle \sqrt{\frac{(1+Tn^2)\tanh(n)}{n}}}\;\cos(nx_i)\cos(nx_j),
\end{equation*}
we have that the above is transformed into the matrix-vector multiplication $K*u_h =\boldsymbol{K} \boldsymbol{u}_h,\,$
where $\boldsymbol{u}_h$ is the vector $[u_1,\ldots,u_N]$ whose entries 
are the discrete solution evaluated at the collocation points.
We can therefore collocate Equation \eqref{eqDiscreteEquation} in the collocation points $x_i$,
and obtain a system of $N$ nonlinear equations
\begin{equation}
	\label{eqVectorEquation}
		W_h(\mu,\boldsymbol{u}_h) = -\mu \,\boldsymbol{u}_h + \boldsymbol{u}_h^2 + \boldsymbol{K}\, \boldsymbol{u}_h = 0.
\end{equation}

\subsection{Choice of parametrisation}
Problem \eqref{eqVectorEquation} requires solving a nonlinear system of equations, written in general from as $F(y)=0$.
This can be done with standard Newton iterations $y^{n+1} = y^n - \left(J_{F}(y^n)\right)^{-1} \, F(y^n)$,
where $J_F$ is the Jacobian of $F$.
Choosing different $F$'s allows to parametrise the problem in different ways, depending on what is most convenient at any given time.
We present here two possible strategies to parametrise and follow the bifurcation branch: One is based on parameter-continuation, while the other is based on the pseudo-arclength method.

\subsubsection{Parameter-continuation approach}
The idea of a parameter-continuation approach consists in choosing a quantity $p$ to be the parameter, it can be for example the speed of the wave, and then in setting $F$ so that a solution to $F(y)=0$ will satisfy \eqref{eqVectorEquation} as well as a constraint linked to the parameter we have chosen.
Once a solution is found, the parameter is updated by a small step $p \leadsto p+h$ and a new solution is computed.
Looking at \eqref{eqVectorEquation}, the most natural choice seems to be
\begin{equation}
	\label{eqSpeedParametrisation}
	F_\mu(u) = W_h(\mu, u_h),
\end{equation}
which corresponds to using the speed as a parametrisation of the branch.
We can picture the branch as a curve plotted in the $(\mu, \zeta)$ plane, where $\zeta$ can be any other quantity used as vertical axis, e.g. the wave height.
Given a fixed speed $\bar{\mu}$ and a corresponding solution $u$ of \eqref{eqSpeedParametrisation}, we modify the speed with a small step $\bar{\mu}+h$ and use the previous solution $u$ as an initial guess for Newton.
The algorithm will then ``move'' vertically from the point $(\bar{\mu}+h, u)$ and converge to a new solution on the branch with speed equal to $\bar{\mu}+h$.
While this is very robust numerically, it clearly breaks down when the curve has a turning point or a vertical tangent, as then the implicit function theorem no longer applies.

Since we already know that nontrivial branches have a vertical tangent at the bifurcation point,
and that turning points may happen, we want to include other types of parametrisations.
Since $\mu$ can no longer be used as a parameter, it needs to be treated as an unknown, and consequently we must include an additional equation in the system.
One idea would be to use the waveheight as a parameter, and we can identify the numerical wave height as $\abs{u_N - u_1}$. We will then choose $F$ to be
\begin{equation}
	\label{eqWhParametrisation}
	F_\textsc{wh}(u,\mu) =
		\begin{pmatrix}
			W_h(\mu,u_h) \\
			u_N - u_1 - \textsc{wh}
		\end{pmatrix}
\end{equation}
where $\textsc{wh}\in\mathbb{R}_+$.

While this is convenient in case of turning points or vertical tangents in the branch, it is based on the assumption that $\abs{u_N - u_1}$ really describes the wave height, i.e. that $u_N$ and $u_1$ are the crest and the trough of the wave.
As we have seen, however, there are cases where the wave can be multimodal, and therefore crests may not be positioned at $u_N$.
See for example Figure \ref{fig1and7d}.
When this happens, this parametrisation will not give any control on the height of the wave, and may make it difficult to accurately follow the branch.

A third option that can be used is to parametrise the curve with the square of the $L^2$-norm of the solution.
This results in a choice of $F$ as
\begin{equation}
	\label{eqL2Parametrisation}
	F_\textsc{l2}(u,\mu) =
	\begin{pmatrix}
		W_h(\mu,u_h) \\[2pt]
		\frac{1}{N}(u_1^2 + u_2^2 + \ldots u_N^2) - \textsc{l2}
	\end{pmatrix}
\end{equation}
where as before $\textsc{l2}\in\mathbb{R}_+$.

Our strategy in the parameter-continuation setting is to perform the first few iterations along the branch using the discrete $L^2$-norm parametrisation,
then switch to \eqref{eqSpeedParametrisation}.
At every step we control the conditioning of the Jacobian of the parametrisation in use,
and when it exceeds a certain tolerance we switch to a different parametrisation.

\subsubsection{Pseudo-arclength continuation}
\label{secPseudoArclength}
Another continuation method that can be used is the pseudo-arclength,
which is a \emph{predictor-corrector} scheme based on the idea that a natural parametrisation for a curve is the arclength.
Let $y = [\mu, u_1, u_2, \ldots , u_N]$. Given a solution $y_n$ on the branch,
we compute the next solution $y_{n+1}$ in three steps: First we compute the tangent vector $z_n \in \mathbb{R}^{n+1}$ at $y_n$ solving
\begin{equation}
	\label{eqTangentVector}
	\begin{cases}
		D_{\mu,u_h}W_h(y_n)\cdot z_n = 0 \\
		z_n\cdot z_{n-1} = \alpha
	\end{cases}
\end{equation}
where $\alpha > 0$. The first of \eqref{eqTangentVector} is the tangency condition,
while the second is used to choose the tangent vector with the correct orientation.

Then, given $z_n$ (properly normalised) we compute $y_{n+1}^p$, \emph{predictor} point to $y_{n+1}$, simply by
\begin{equation}
	\label{eqPredictorPoint}
	y_{n+1}^p = y_n + h\,z_n.
\end{equation}

Finally, the new point $y_{n+1}$ is found by projecting $y_{n+1}^p$ onto the branch in a direction perpendicular to $z_n$.
That is, we obtain $y_{n+1}$ by solving
\begin{equation}
	\label{eqProjection}
	\begin{cases}
		W_h(y_{n+1})=0 \\
		(y_{n+1} - y_{n+1}^p)\cdot z_n = 0
	\end{cases}
\end{equation} 
which is the \emph{corrector} step of the method.

This method is surprisingly robust, and enables us to easily follow the branch even in presence of turning points.
It is clear, however, that it requires an initial guess for the first tangent vector $z_0$.
The last $N$ components of $z_0$ can be chosen, according to what was said in Section \ref{secBifurcationFormulas},
as $\cos(kx_i), i=1,\ldots,N$.
For the first component the optimal choice would be to use the information coming from \eqref{eqExpansionFormulas_c};
however as we have noted that expansion is not always valid.
In order to circumvent this problem and obtain information on the ``direction'' for the speed,
we decided to first use the parametrisation \eqref{eqL2Parametrisation}, which returns some value $\tilde{\mu}_1$ as a solution,
and then simply set the first component in $z_0$ as $\sign(\tilde{\mu}_1 - \mu^*)$.

\section{Numerical Results}
\label{secResults}

We present in this section the numerical results we obtained applying the scheme presented earlier.
All results have been obtained employing the pseudo-arclength parametrisation
described in Section \ref{secPseudoArclength}.
For notational simplicity we will refer to the branch obtained for $k=1$ 
as the {\it main} branch even in the presence of multi-modal waves.

The computed profiles have been tested in a discrete time integrator
in order to ascertain their validity as numerical solutions of the Whitham problem.
To this end, a fully discrete time dependent collocation scheme was developed which is similar
to the scheme used in \cite{Ehrnstrom2013gbf}. 
While a detailed discussion of the time integration scheme and corresponding results 
is beyond the scope of this work, we note that well posedness of a class of nonlocal equations
was proved in \cite{LPS}, and convergence of spectral collocation projections
of similar nonlocal equations was proved in \cite{HK,PD}, so that this discussion is therefore omitted here.

\subsection{General branches}

\begin{figure}[t]
	\centering
	
	\subfloat[$T=0$]{
		\includegraphics[width=0.39\linewidth]{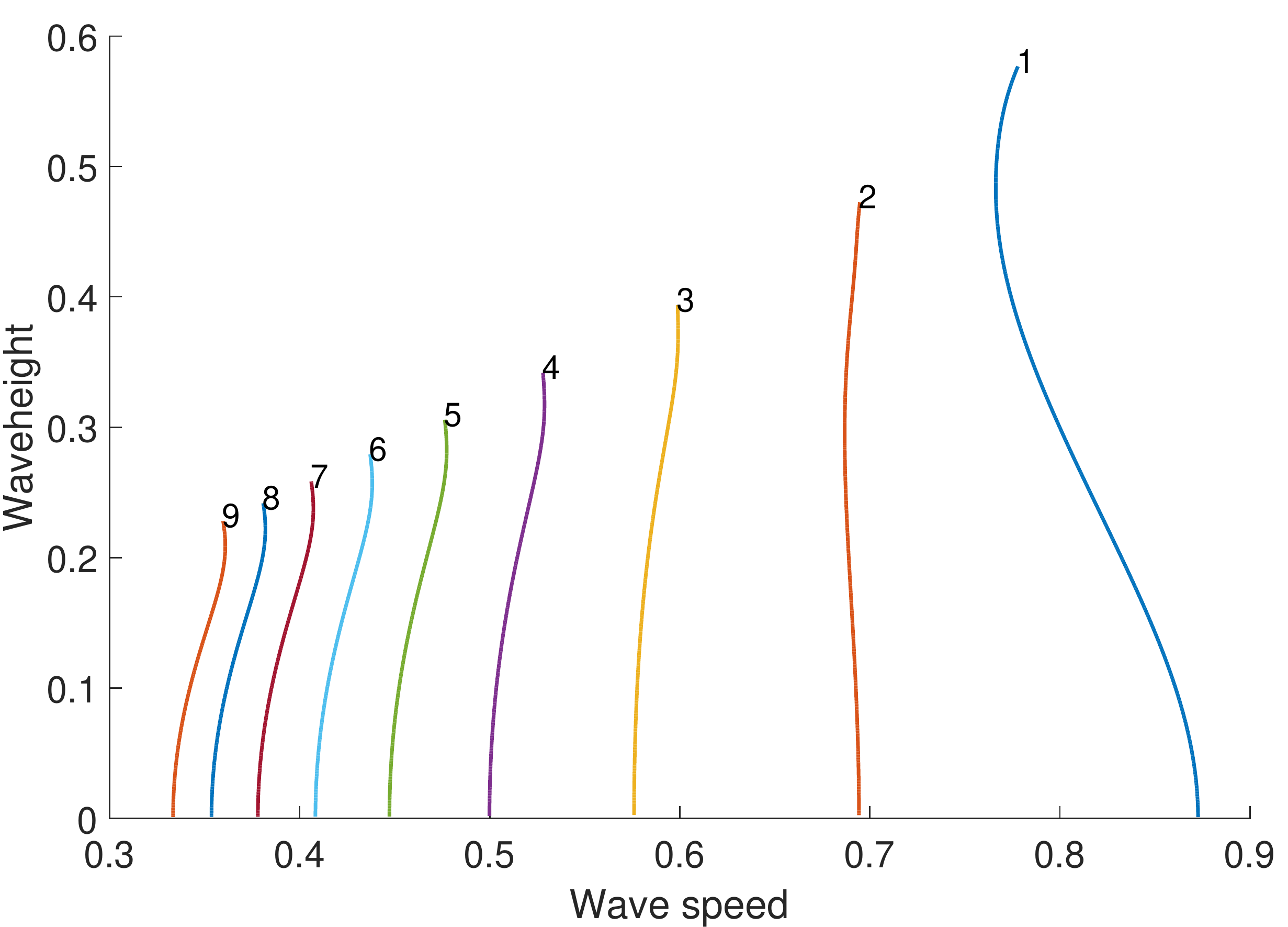}
		\label{figAll0}
	}
	\subfloat[$T=0.1$]{
		\includegraphics[width=0.39\linewidth]{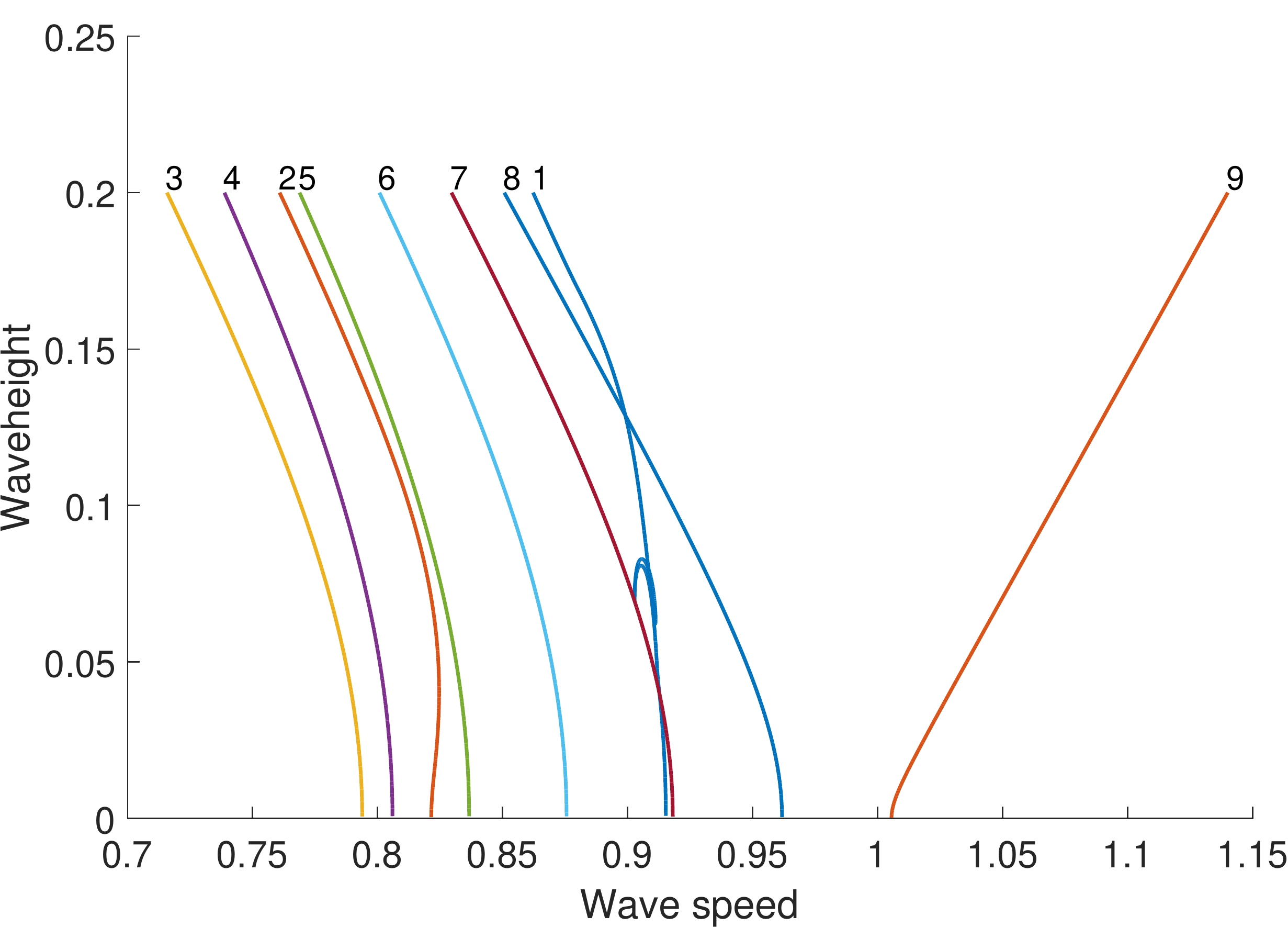}
		\label{figAll01}
	}
	
	\subfloat[$T=0.2$]{
		\includegraphics[width=0.39\linewidth]{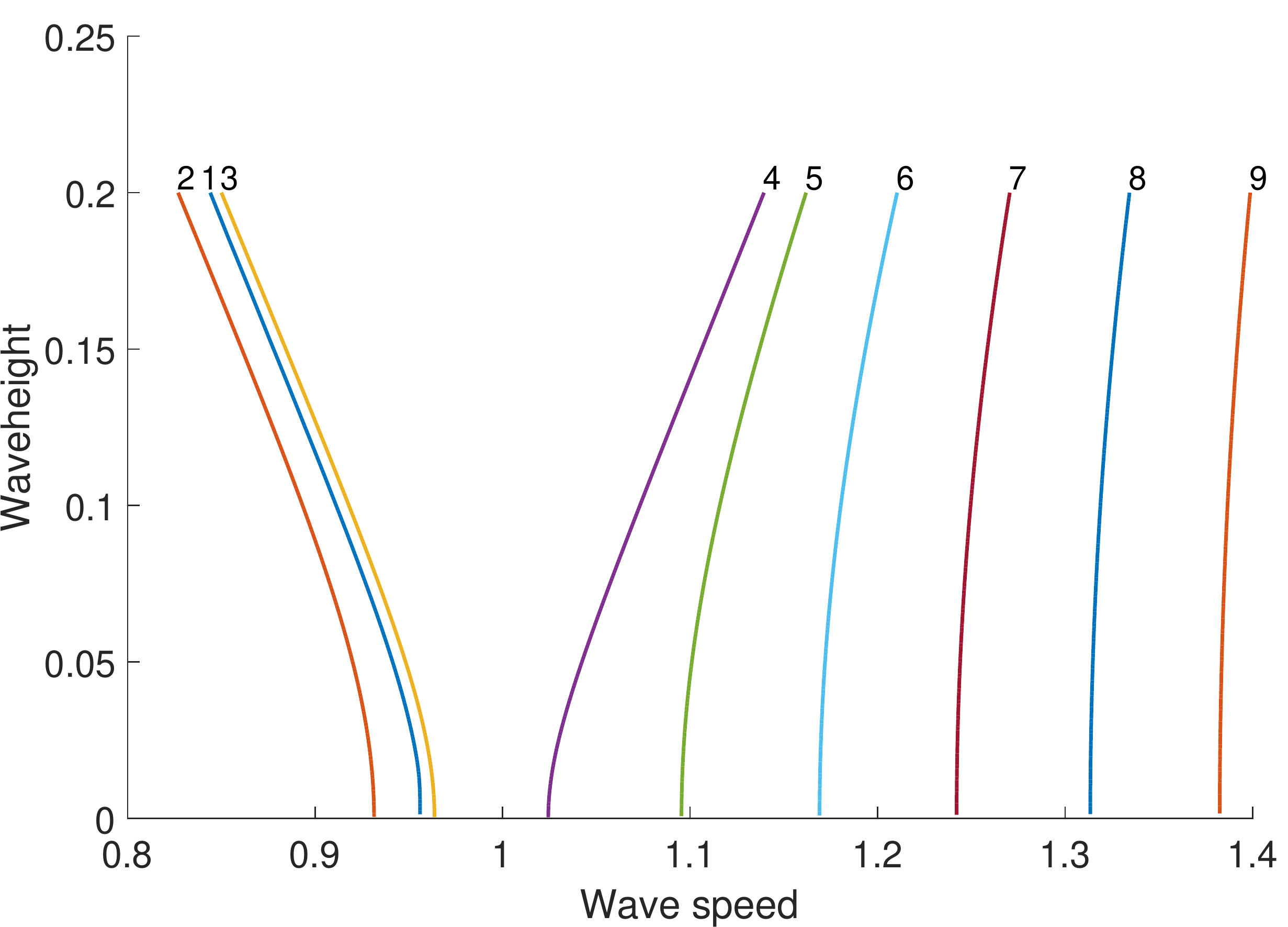}
		\label{figAll02}
	}
	\subfloat[$T=0.333$]{
		\includegraphics[width=0.39\linewidth]{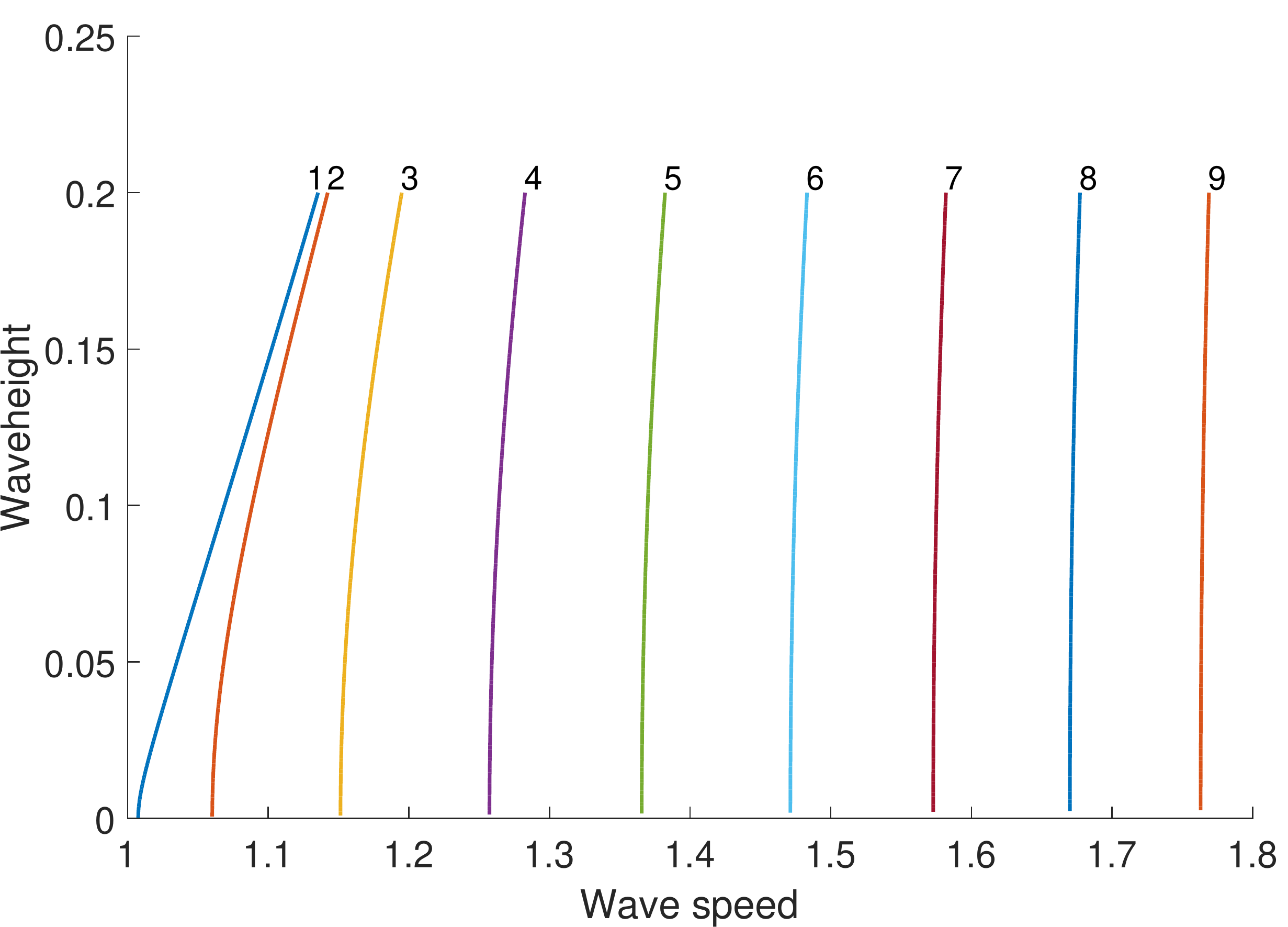}
		\label{figAll0333}
	}
	
	\caption{\label{figAllBranches} 
Branches of solutions for $k=1,\ldots,9$ for different values of the capillarity parameter $T$. 
The value of $k$ is indicated above the corresponding branch.
In panel (b), it can be seen that the branch $k=1$ crosses both the branch $k=7$ and the branch $k=8$.
}
\end{figure}

Figure \ref{figAllBranches} presents the plots of the branches for $k=1,\ldots,9$, for the same values of the capillarity parameter $T$ as considered in Figure \ref{figBifurcationSpeed}.

\begin{figure}[htb]
	\centering
	
	\subfloat[$T=0.1$]{
		\includegraphics[width=0.39\linewidth]{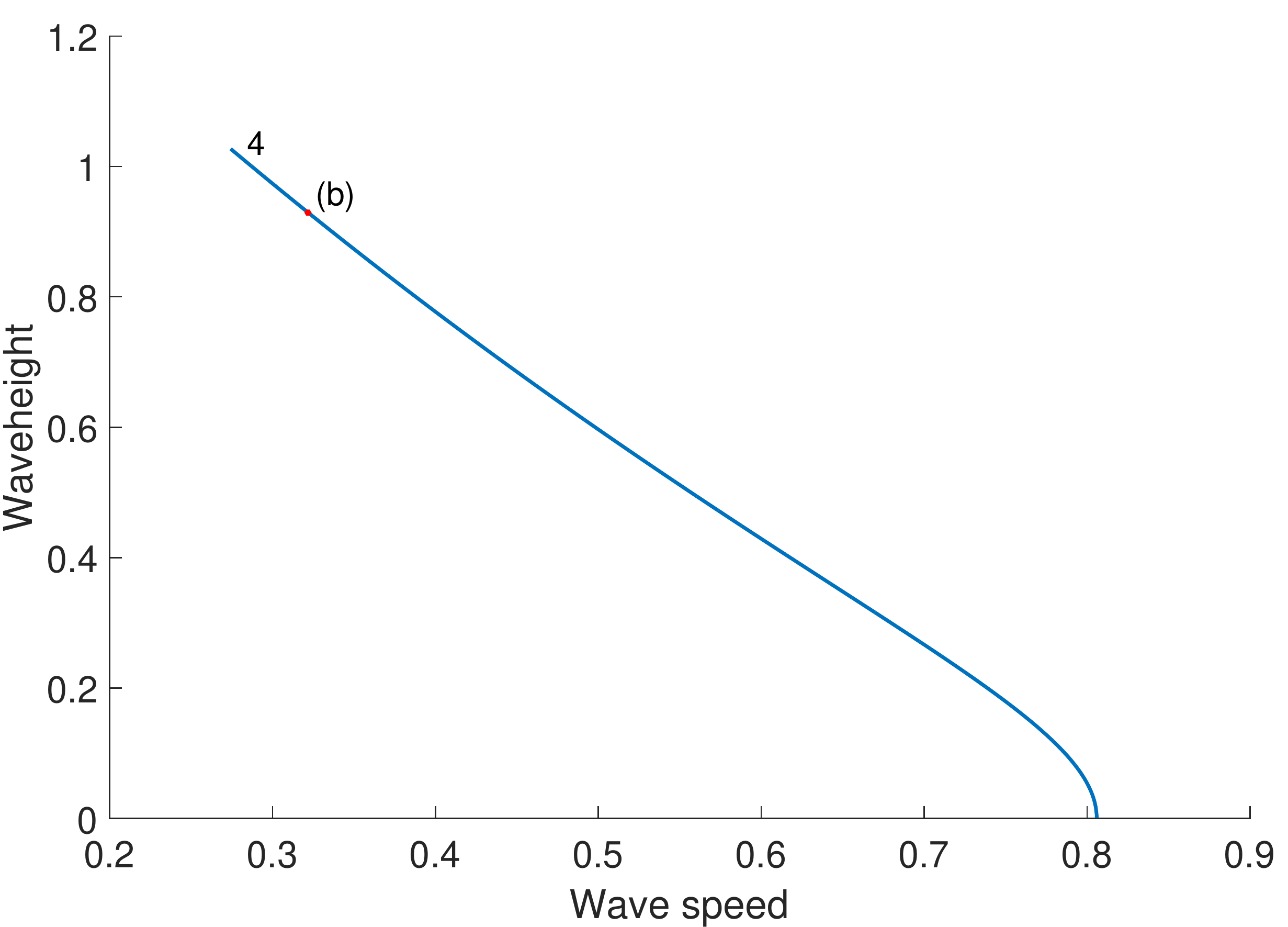}
		\label{figAlmostCuspedWavea}
	}
	\subfloat[]{
		\includegraphics[width=0.39\linewidth]{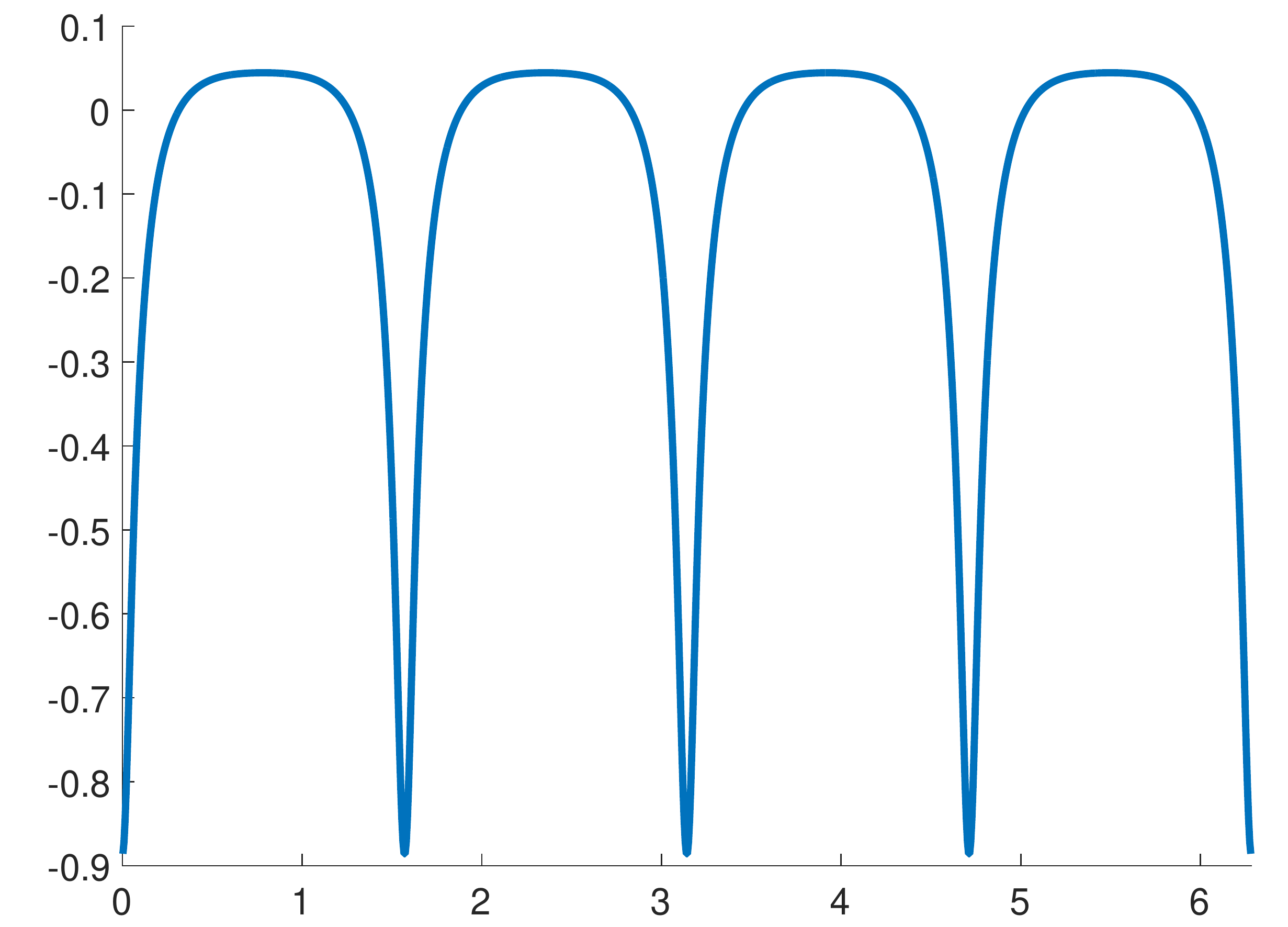}
		\label{figAlmostCuspedWaveb}
	}
	
	\caption{\label{figAlmostCuspedWave}
Bifurcation branch for Bond number $T=0.1$ and wave number $k=4$. The bifurcation point is
$\mu^* \approx 0.806$. Panel (b) displays a very steep wave of waveheight close to $1$.
}
\end{figure}

Note that in the purely gravitational case $T=0$, Figure \ref{figAll0}, 
we took advantage of the known theoretical result stating that $u\leqslant \mu/2$.
To the best of our knowledge no similar result is available for the capillary case, hence in Figures \ref{figAll01}, \ref{figAll02}, and \ref{figAll0333} we are showing the branches up to the wave height value of $0.2$, 
in order to keep the plots readable.
It is important to note, however, that the code can continue the branches also after those heights, and in particular we are able to continue the branches well over heights of 1.
In several cases we tested the highest computed profiles in the time integrator and let the profile evolve for several periods; all tested waves resulted to be orbitally stable.
However, these waves may still feature modulational instability, such as discussed in \cite{HurJohnson15a,Sanford}
for the purely gravitational Whitham equation and a more general class of equations in \cite{BHJ2016}.
We also briefly note here that waves high up in the branches may have very steep profiles, 
which in turn makes the time evolution error very sensitive to the stepsize used.
We present one such example in Figure \ref{figAlmostCuspedWave}.

To the naked eye the plot of the profile can appear so steep that it seems to almost develop cusps of depression.
From the theory it is clear that any solution of the Whitham problem has to be smooth, in particular $C^\infty$, so cusps cannot really develop, but this may be an indication of a possible blow-up in the derivative.

Going back to Figure \ref{figAllBranches} we see that, as expected, the bifurcation speed of the branches increases with $T$ and for $T>\frac{1}{3}$, $\mu^*(k)>1 \, \forall k$.
We can also see that the branches bifurcate in the direction of increasing or decreasing velocities in accordance with Section \ref{secConcavity} (see also Figure \ref{figConcavity}).
Moreover, we note that turning points are present also in the capillary case: See for example the branch for $k=2$ in Figure \ref{figAll01}.

Section \ref{secConnectingBranches} contains a more detailed discussion of
the complex interaction happening between the main branch and the branch for $k=7$ 
that can be seen in Figure \ref{figAll01}.

\subsection{Two-dimensional Bifurcation}
\label{sec2Dim}

This section is devoted to the cases where $T$ is chosen as in \eqref{eqTForCollision}.
For these values we know from Section \ref{secBifurcationFormulas} that the bifurcation kernel is two-dimensional and the analytical expansions with the coefficients given in \ref{secExpansionCoeffs} may no longer be valid.
Also, as will be proved in future work, the two-dimensional bifurcation kernel 
leads to the existence of two-dimensional \emph{sheets} of small amplitude solutions.
Our code, however, is currently capable of following only \emph{branches} of solutions. 
In the case of a two-dimensional kernel, this corresponds to following the intersection 
curve between the sheet of solutions and  the plane $T=\text{const}$.

\begin{figure}[tb]
	\centering
	
	\subfloat[$T=T(1,7) \approx 0.09918$]{
		\includegraphics[width=0.39\linewidth]{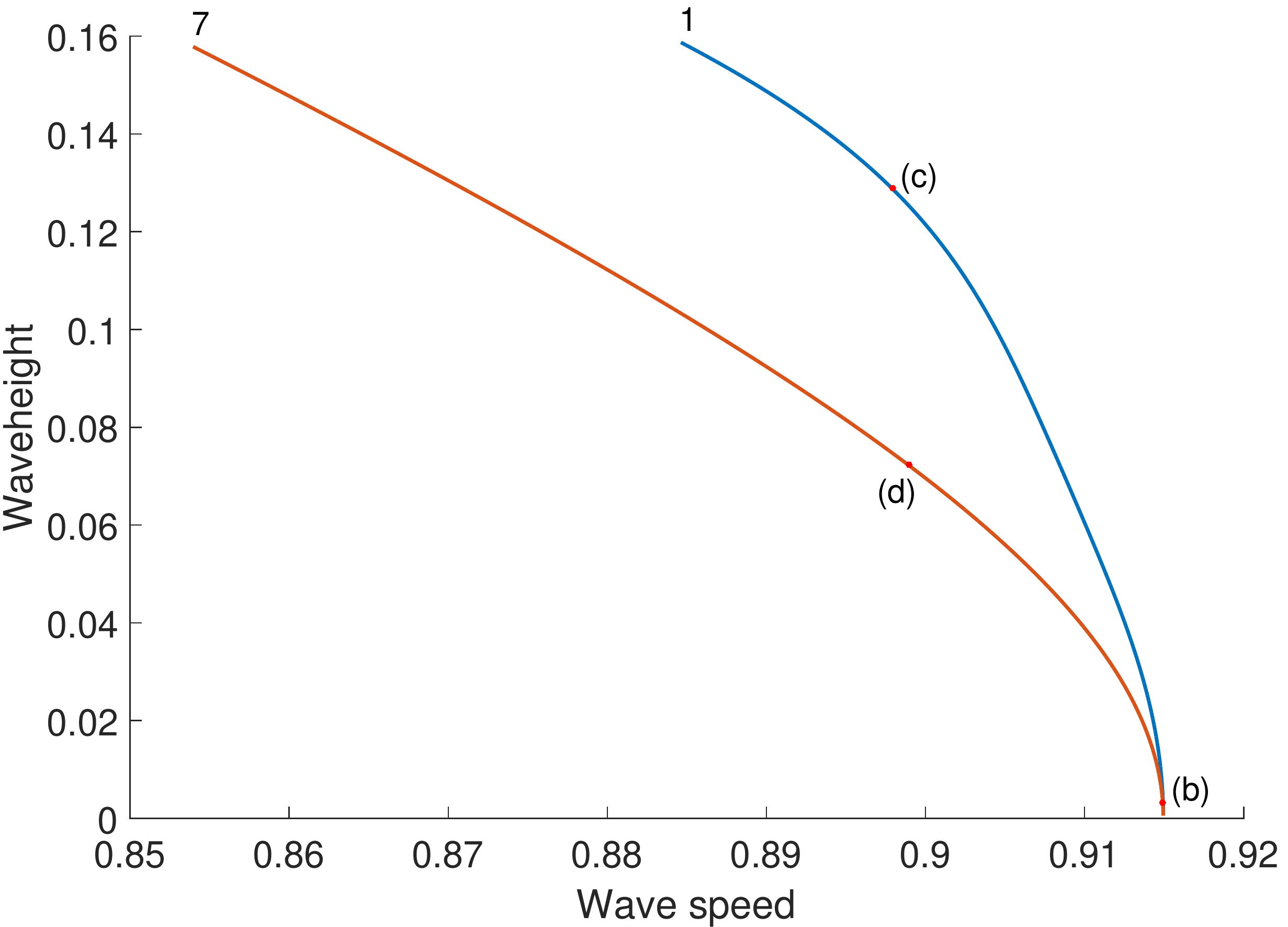}
		\label{fig2Dim-1-7a}
	}
	\subfloat[]{
		\includegraphics[width=0.39\linewidth]{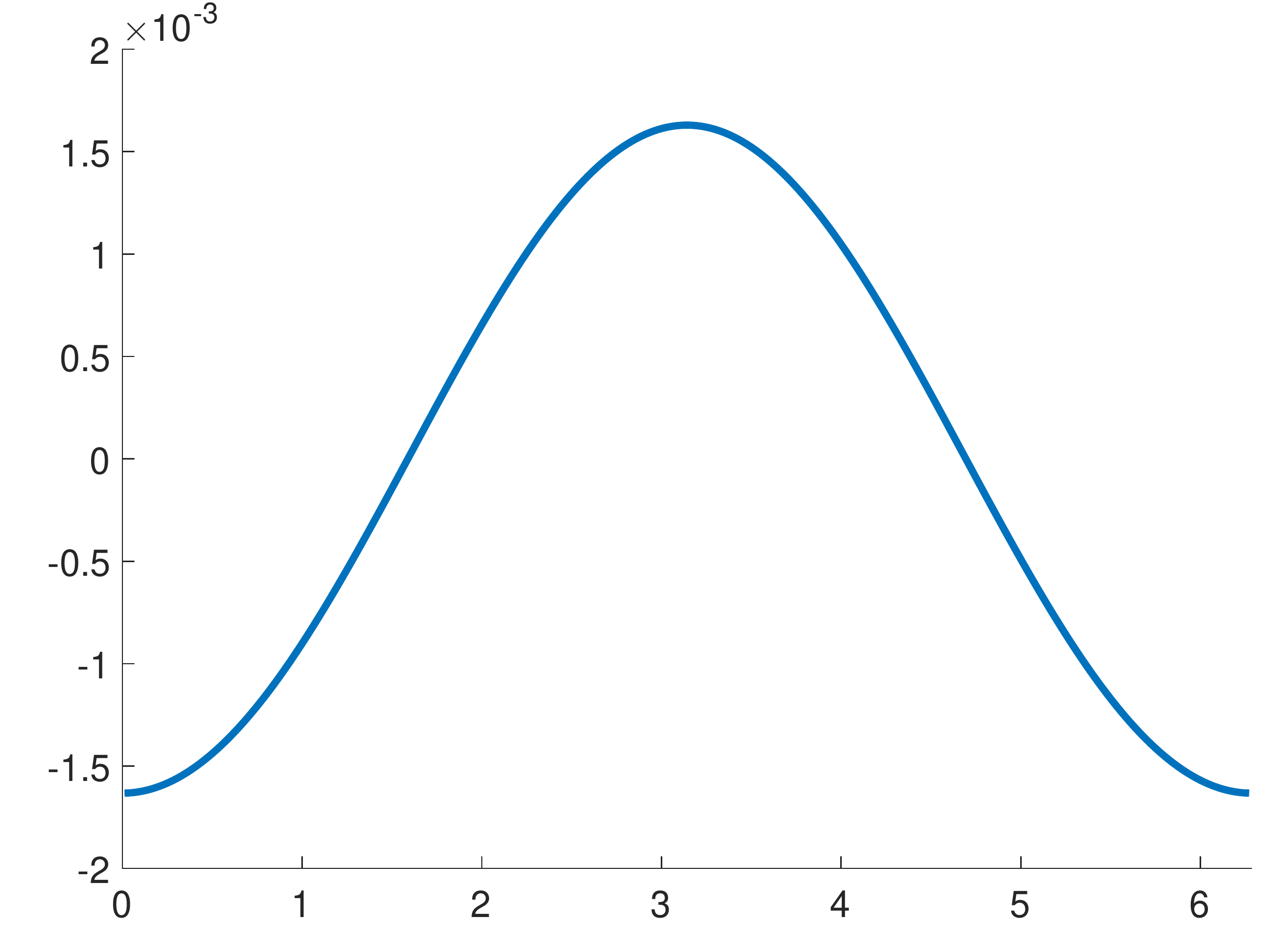}
		\label{fig2Dim-1-7b}
	}
	
	\subfloat[]{
		\includegraphics[width=0.39\linewidth]{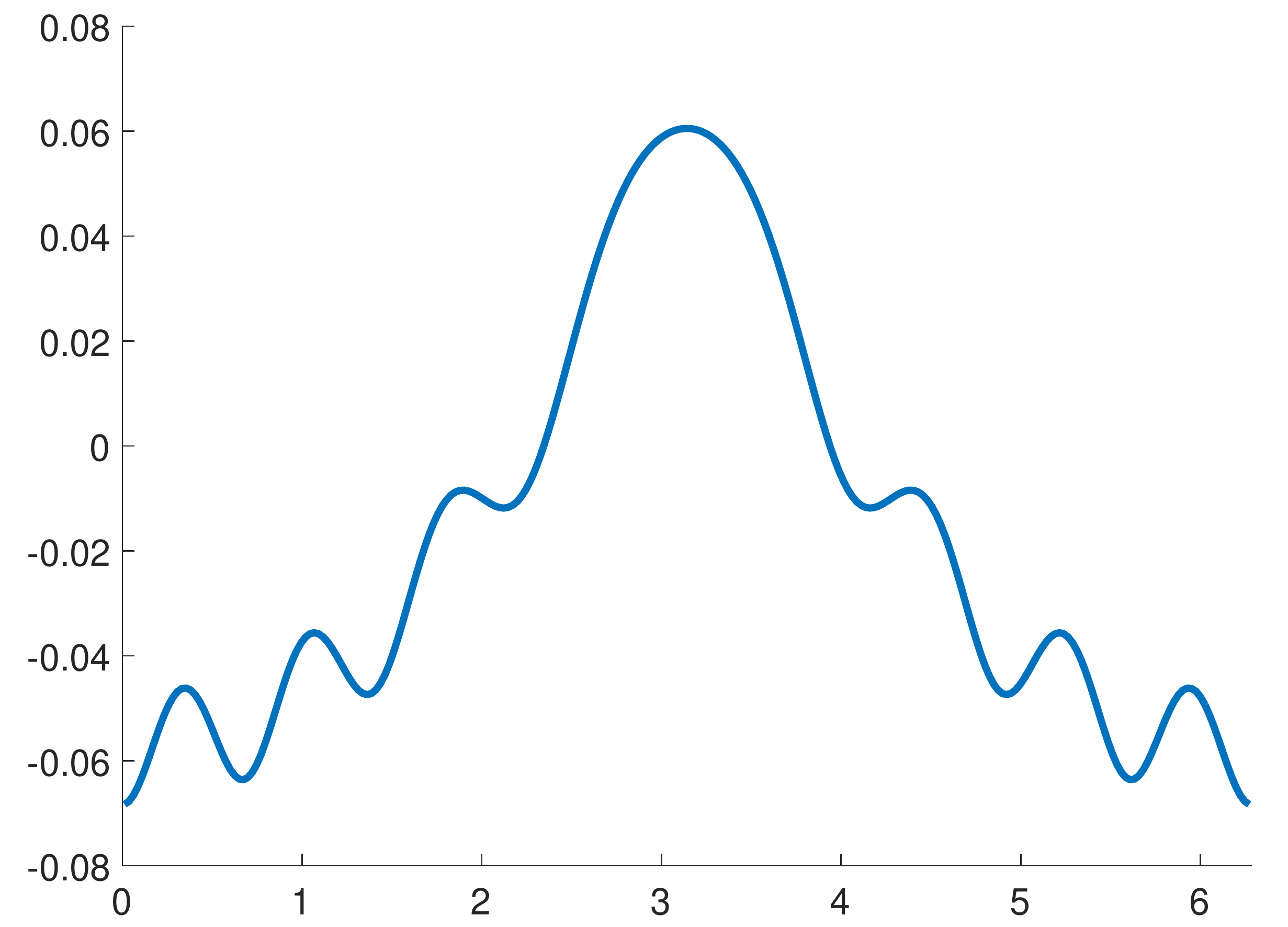}
		\label{fig2Dim-1-7c}
	}
	\subfloat[]{
		\includegraphics[width=0.39\linewidth]{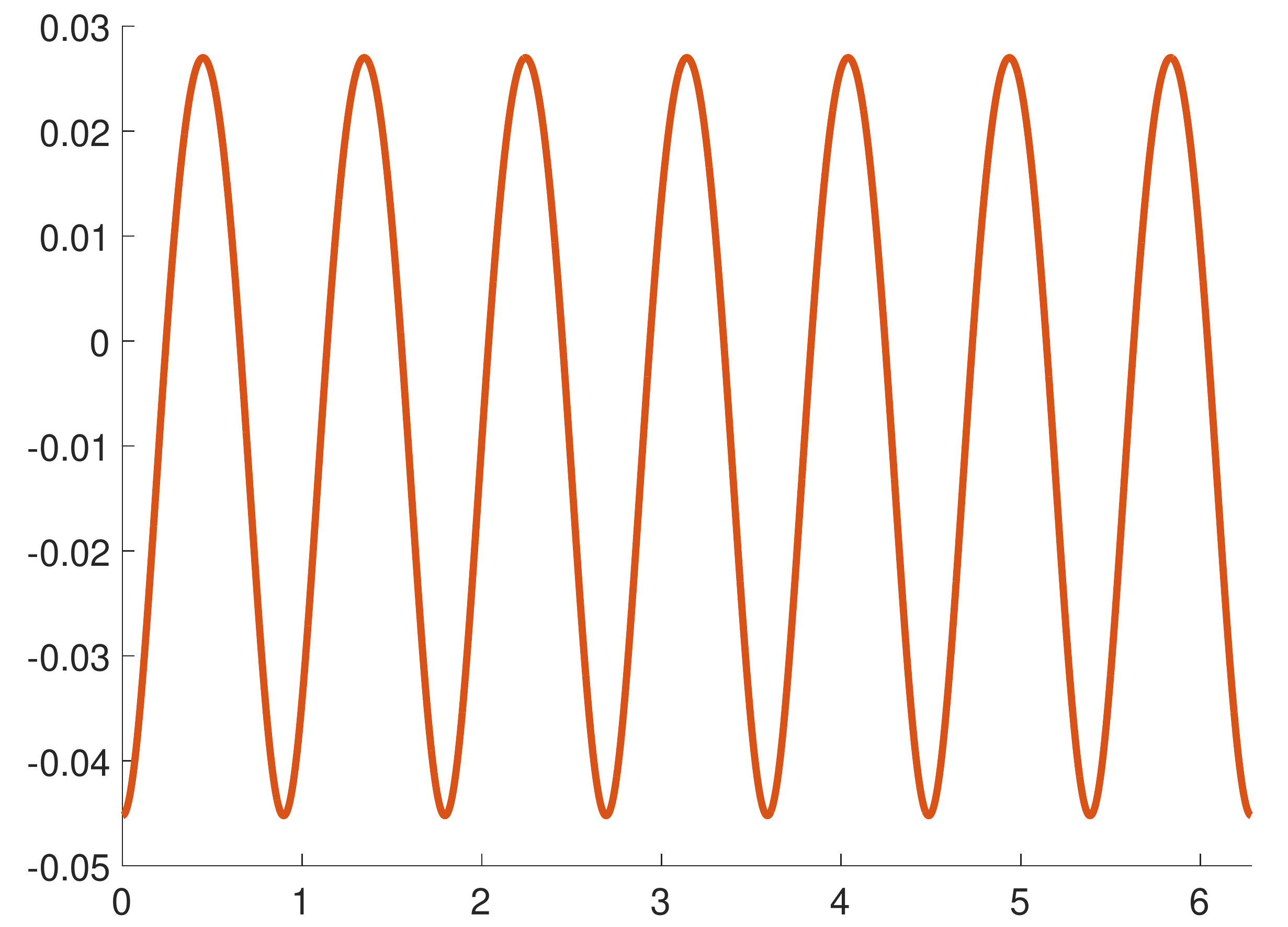}
		\label{fig2Dim-1-7d}
	}
	
	\caption{\label{fig2Dim-1-7}
Two bifurcation branches originating from the same bifurcation point $\mu^* \approx 0.915$.
Here $T$ is given by $T(1,7)$ according to formula \eqref{eqTForCollision}. Panel (a) shows the two 
bifurcation branches. Panels (b) and (c) show wave profiles on the upper (blue) bifurcation curve, 
and panel (c) shows a wave profile on the lower (red) bifurcation curve.
}
\end{figure}

Figure \ref{fig2Dim-1-7} shows the plot for the branches for  $k_1=1$ and $k_2=7$ when {$T = T(1,7) \approx 0.09918$}.
Note that the profiles of the waves at the points labelled (b), (c), and (d) are shown in the corresponding subfigures.
In this case the bifurcation kernel is spanned by $\{\cos(x), \cos(7x)\}$ and the branches bifurcate from the same point as expected.
As we can see the main branch contains waves with mixed wavenumbers: At the beginning (Figure \ref{fig2Dim-1-7b}) waves have simple cosine-like profiles, but further up the branch (Figure \ref{fig2Dim-1-7c}) the influence from the $\cos(7x)$ component becomes more pronounced and they develop 7 crests.
This change happens somewhat rapidly in the lower part of the branch, while in the higher part the profiles seem to stabilise to a mix of $\cos(x)$ and $\cos(7x)$, and little change in shape is observed between waves even over great distances in the branch.
The waves in the $k_2$ branch, on the other hand, maintain a pure $\cos(7x)$-profile throughout the branch.

\begin{figure}[htb]
	\centering
	\subfloat[$T=T(1,2) \approx 0.23968$]{
		\includegraphics[width=0.39\linewidth]{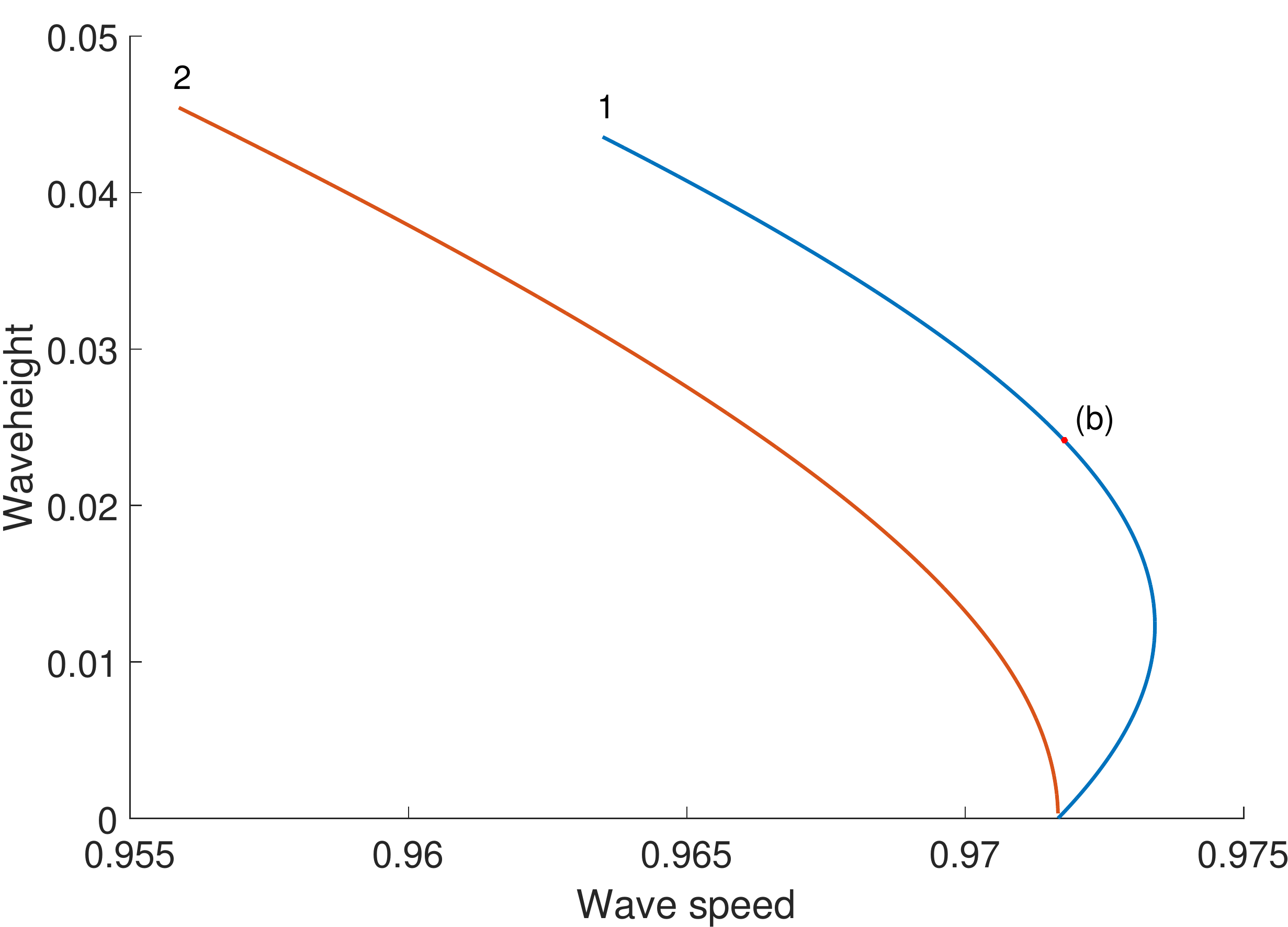}
		\label{fig2Dim-1-2a}	}
	\subfloat[]{
		\includegraphics[width=0.39\linewidth]{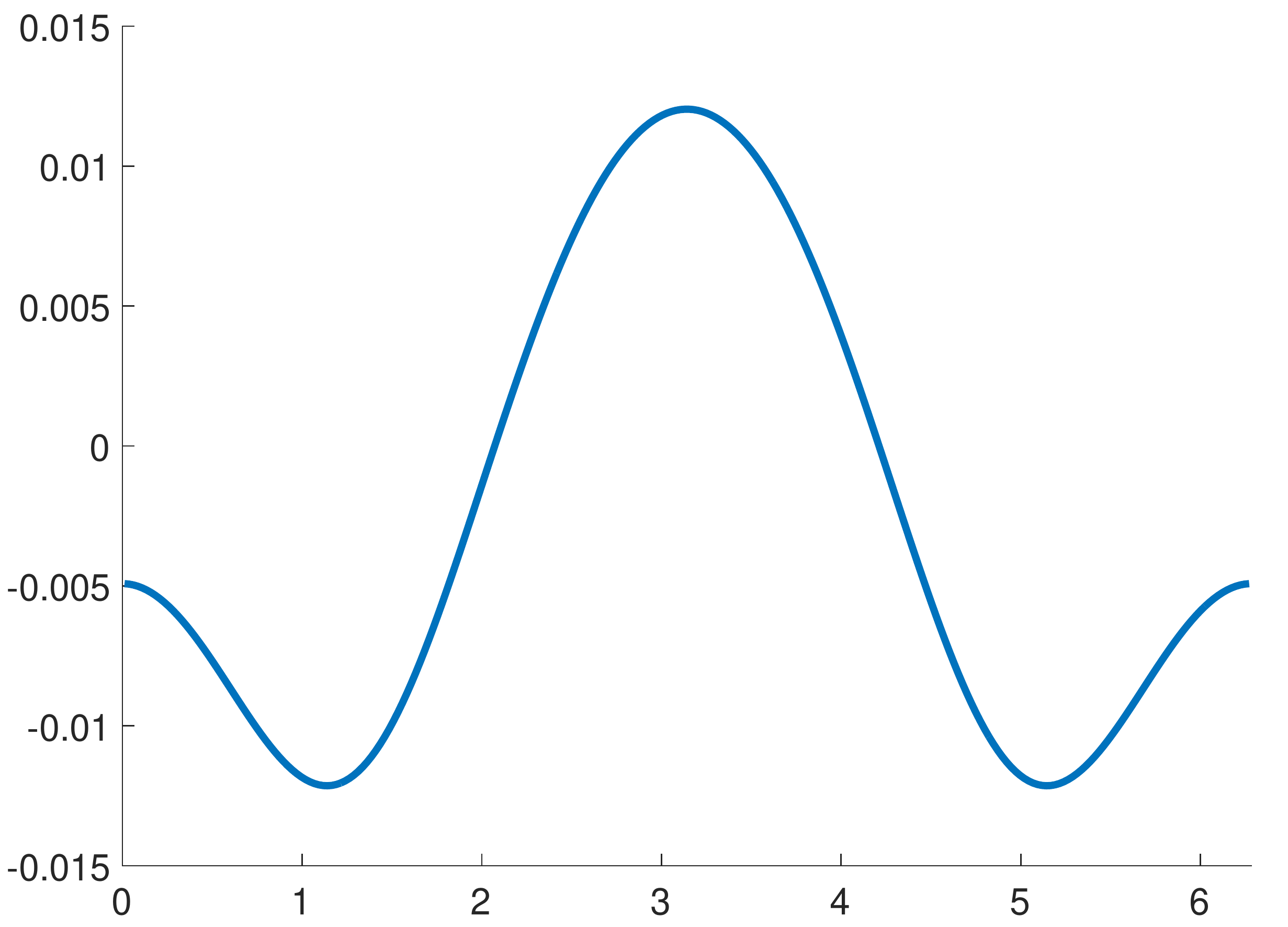}
		\label{fig2Dim-1-2b}	}
	\caption{\label{fig2Dim-1-2}
A Pair of bifurcation branches originating from the bifurcation point $\mu^* \approx 0.972$. The Bond number $T$ is given by $T(1,2)$ according to 
formula \eqref{eqTForCollision}. Panel (a) shows the two bifurcation branches. 
Panel (b) shows a wave profile on the upper (blue) bifurcation curve.
}
\end{figure}

Changing $T$ with the help of \eqref{eqTForCollision} we can produce two-dimensional kernels containing any $k_1$ and $k_2$: Figure \ref{fig2Dim-1-2} shows a case similar to the above for the wavenumbers $k_1=1$ and $k_2=2$.
Note that since now $k_2 = 2\,k_1$, the expansion formulae with the coefficients written in Section \ref{secExpansionCoeffs} are no longer valid, and in particular we see that the $k_1$-branch does not have a vertical tangent at the bifurcation point.
As in the previous case, the main branch contains mixed waves: In the lower part the principal mode is $\cos(x)$, while as one follows the branch the contribution from the $\cos(2x)$ mode becomes noticeable and the profile develops two crests.
The profile of waves in the $k_2$ branch, instead, is not affected by the lower $k_1$-mode and remains of the form $\cos(2x)$.
This fact is clear from a functional-analytical point of view since $C_{2\pi/2}$, the space of continuous, $2\pi/2$ periodic functions, is a subset of $C_{2\pi}$.

\begin{figure}[htb]
	\centering
	\subfloat[$T=T(3,4) \approx  0.08086$]{
		\includegraphics[width=0.39\linewidth]{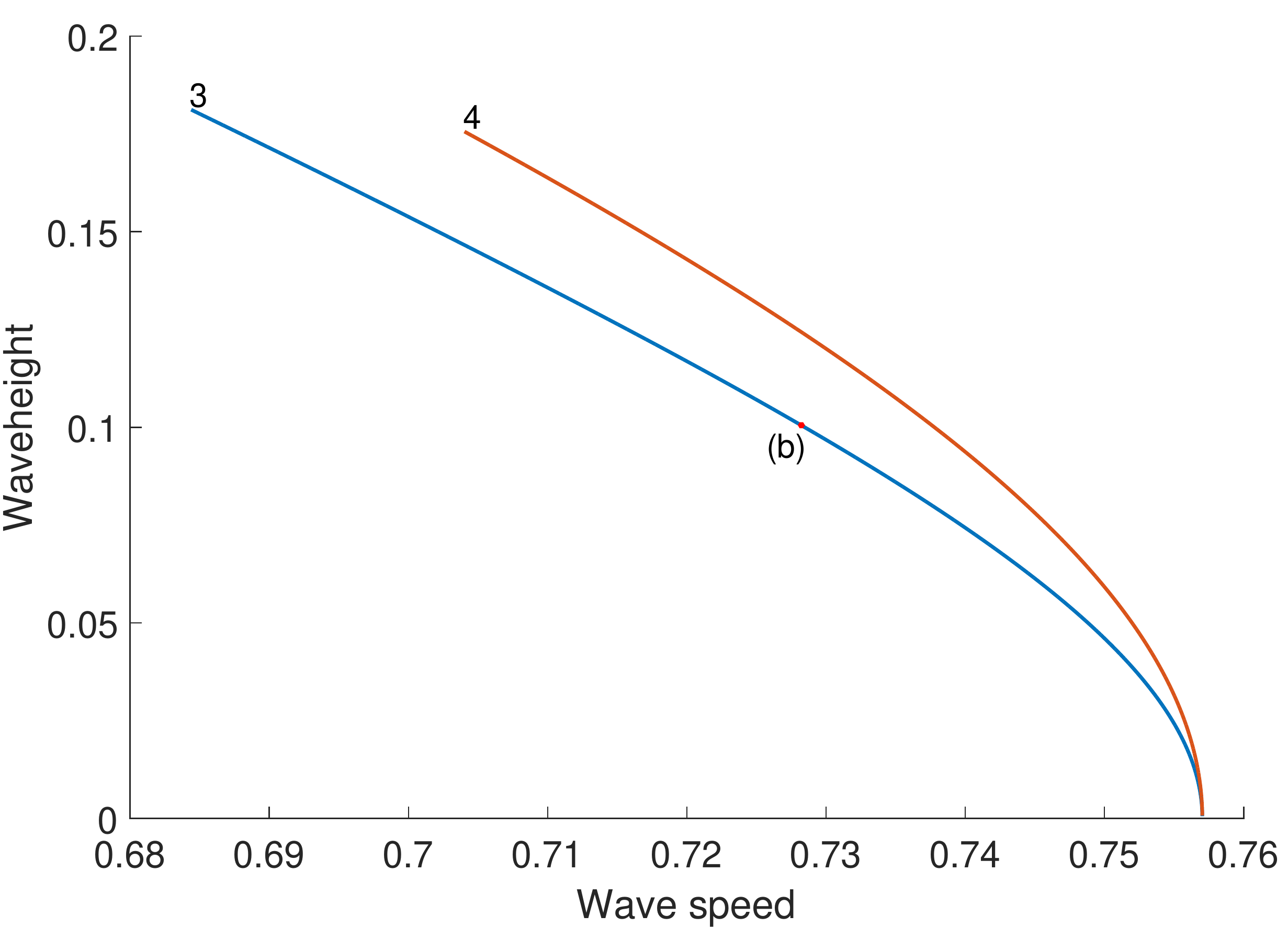}
		\label{fig2Dim-3-4a}
	}
	\subfloat[]{
		\includegraphics[width=0.39\linewidth]{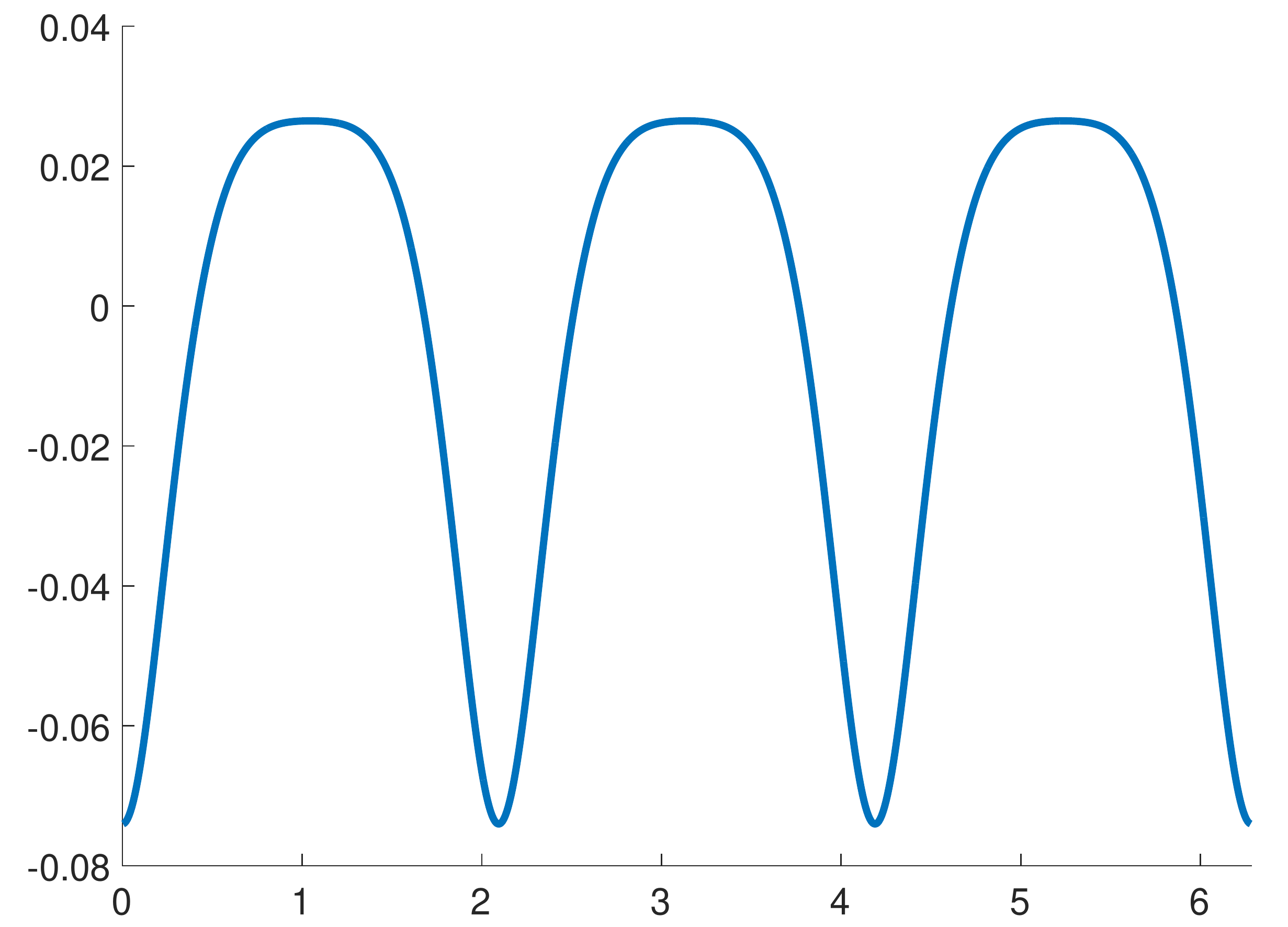}
		\label{fig2Dim-3-4b}
	}
		\caption{\label{fig2Dim-3-4}
A Pair of bifurcation branches originating from the bifurcation point $\mu^* \approx 0.757$.
The Bond number $T$ is given by $T(3,4)$ according to formula \eqref{eqTForCollision}. 
Panel (a) shows the two bifurcation branches. 
Panel (b) shows a wave profile on the lower (blue) bifurcation curve.}
\end{figure}

More generally, if $k_2=a\,k_1$ for some $a\in\mathbb{N}$, 
then $C_{2\pi/k_2} \subsetneq C_{2\pi/k_1}$ and the $k_1$-branch will contain solutions with components mixing the wavenumbers $k_1$ and $k_2$.
If instead $k_2$ is not an integer multiple of $k_1$, {${C_{2\pi/k_2} \nsubseteq C_{2\pi/k_1}}$} and therefore the $k_1$ branch will not contain any component with period $2\pi/k_2$: See for example Figure \ref{fig2Dim-3-4}.
The numerical tests show that it is still possible for the $k_1$-branch to include period-halving components, which will lead to the formation of two new crests in place of the original ones.
Looking at the coefficients in Section \ref{secExpansionCoeffs} it is clear that the height on the branch where this will happen is proportional to $m(k_1)-m(2k_1)$, but anyway there will not be components with pure $k_2$ wavenumbers.

\subsection{Connecting branches}
\label{secConnectingBranches}

In this section we explore in more detail the cases where the $k_1$ branch actually connects to the $k_2$ one.
A first example could already be seen in Figure \ref{figAll01}, where the main branch connects to the $k=7$ branch.
We will look at this case in detail, and briefly present other cases later.

\begin{figure}[!htb]
	\centering
	
	\subfloat[$T=0.1$]{
		\includegraphics[width=0.39\linewidth]{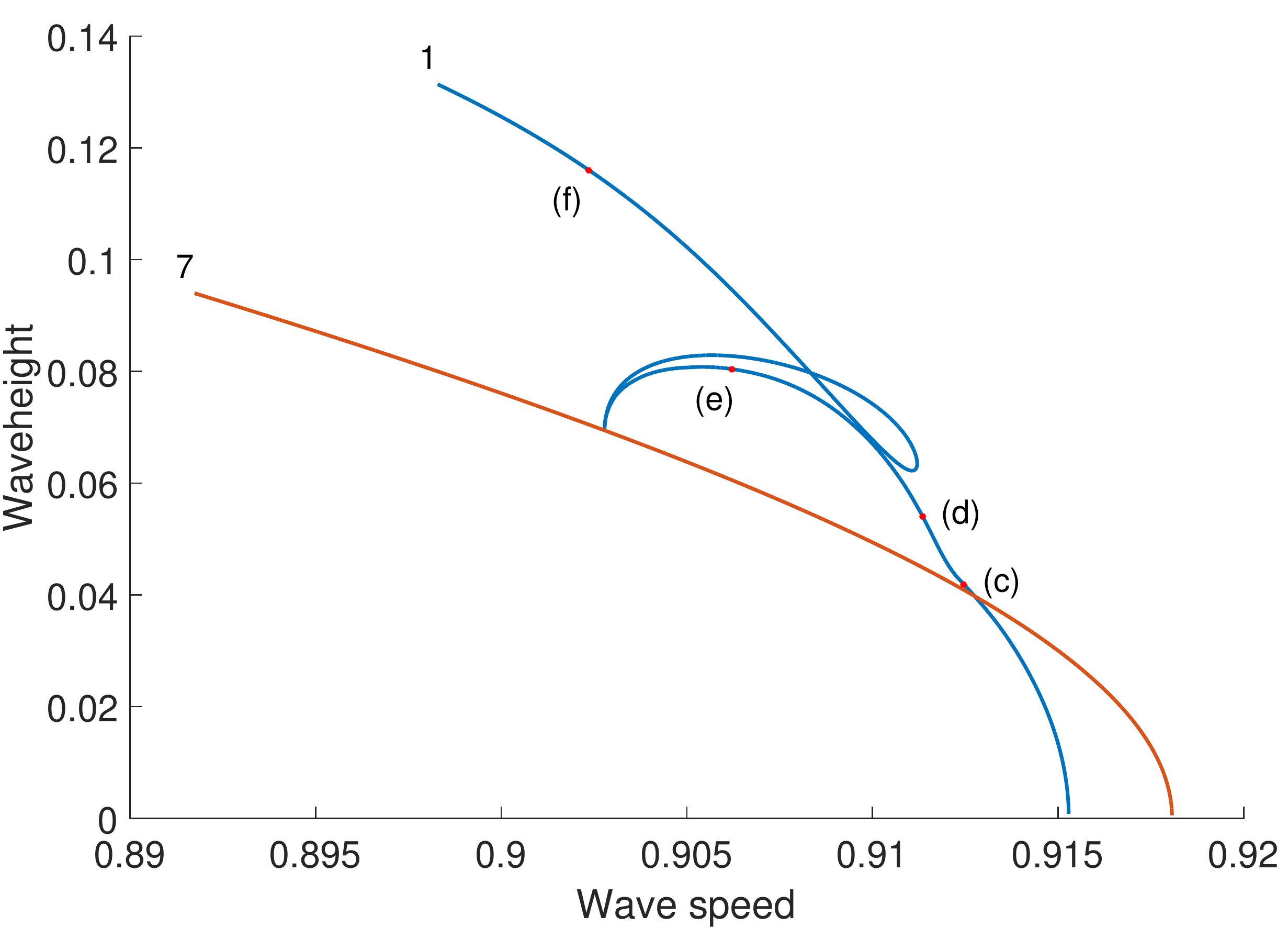}
		\label{fig1and7a}
	}
	\subfloat[]{
		\includegraphics[width=0.39\linewidth]{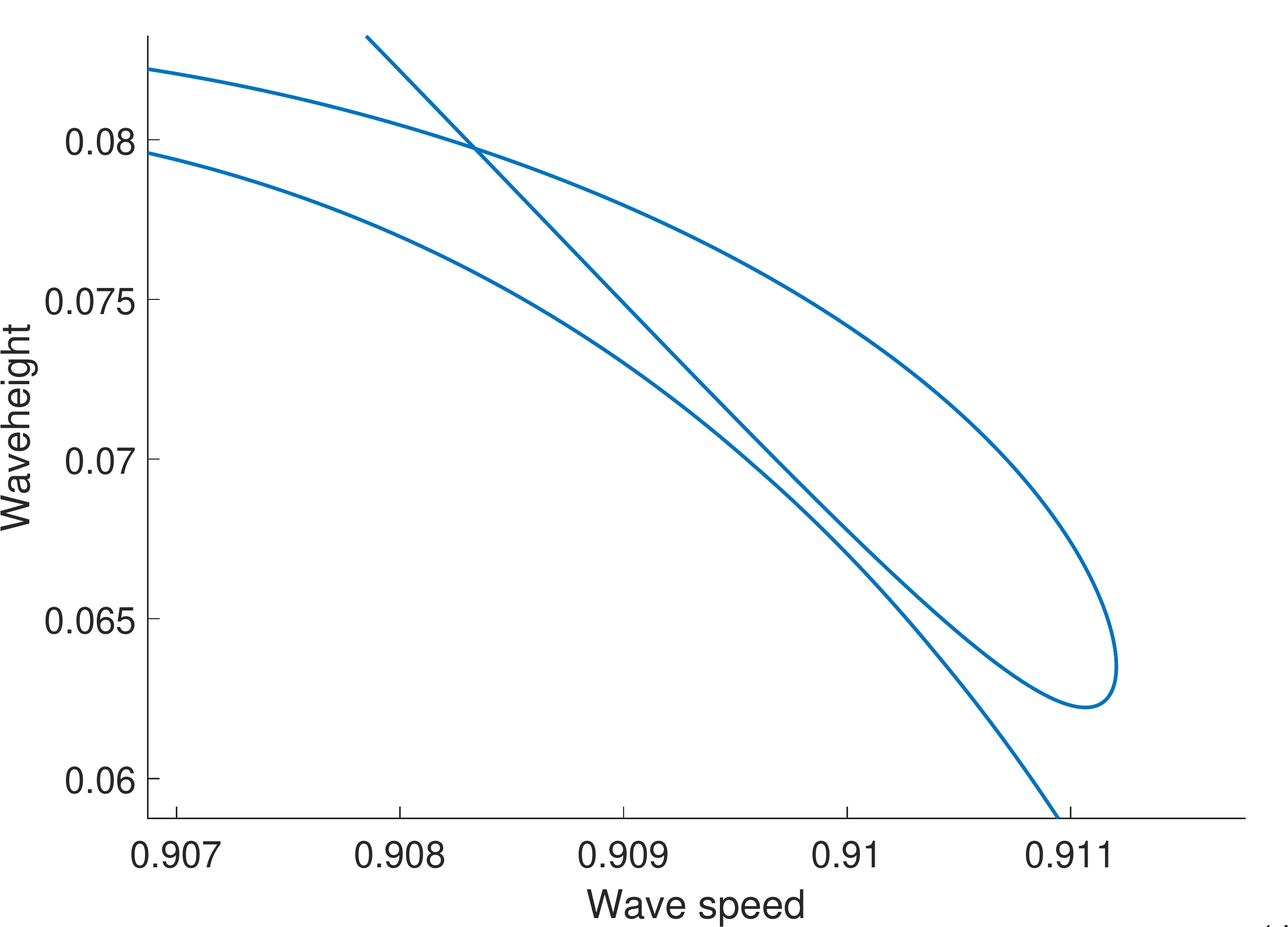}
		\label{fig1and7b}
	}
	
	\subfloat[]{
		\includegraphics[width=0.39\linewidth]{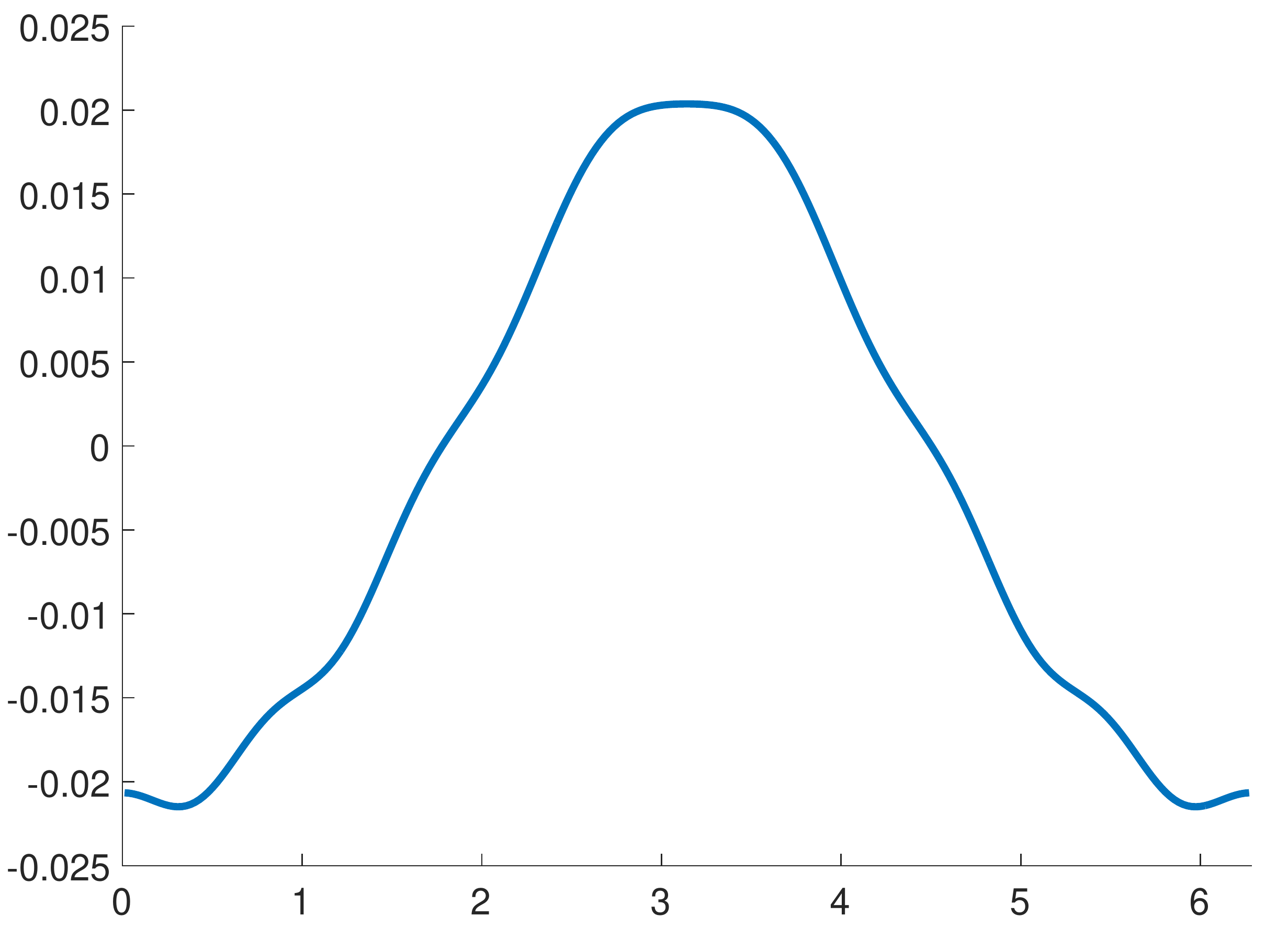}
		\label{fig1and7c}
	}
	\subfloat[]{
		\includegraphics[width=0.39\linewidth]{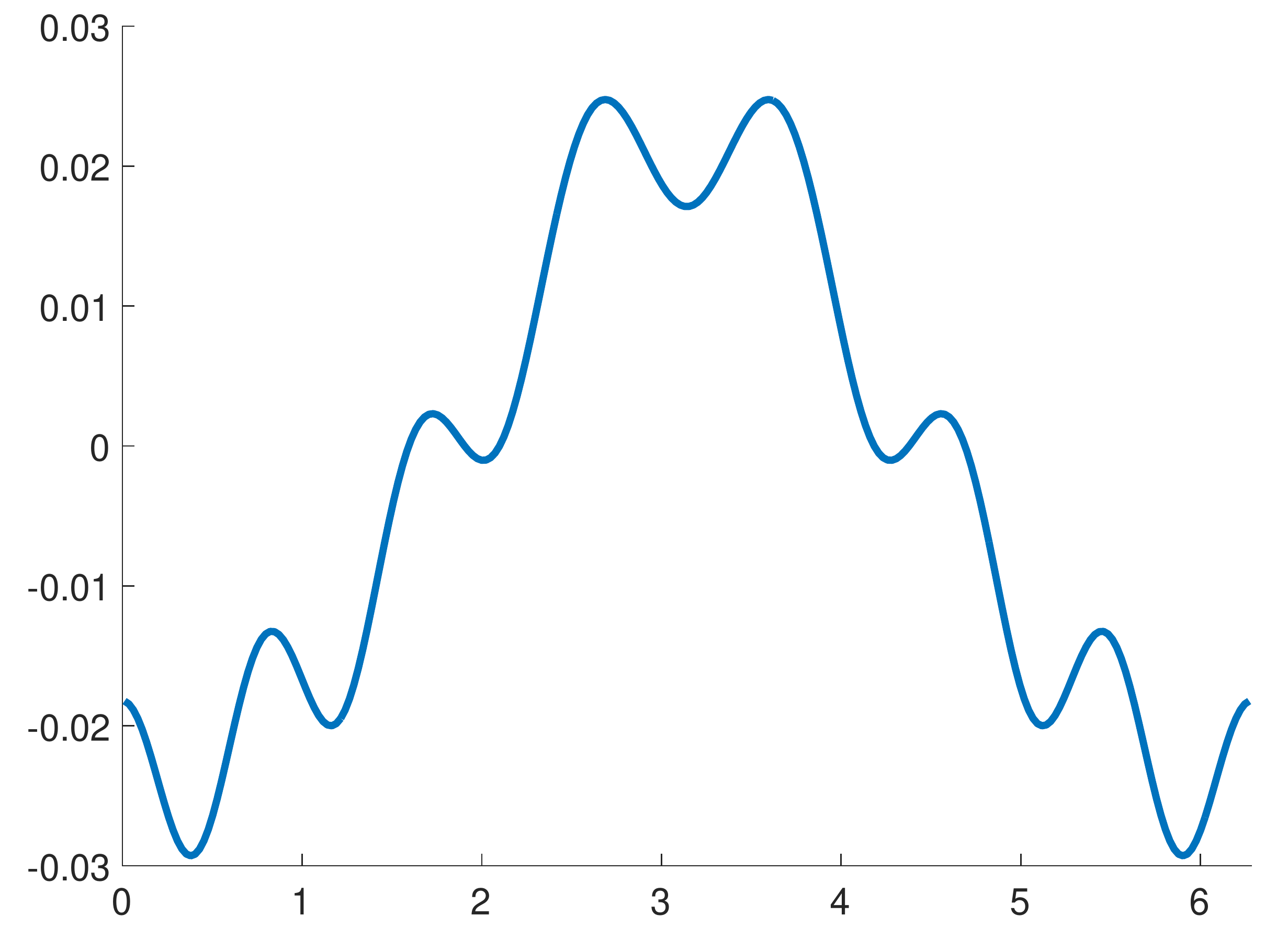}
		\label{fig1and7d}
	}
	
	\subfloat[]{
		\includegraphics[width=0.39\linewidth]{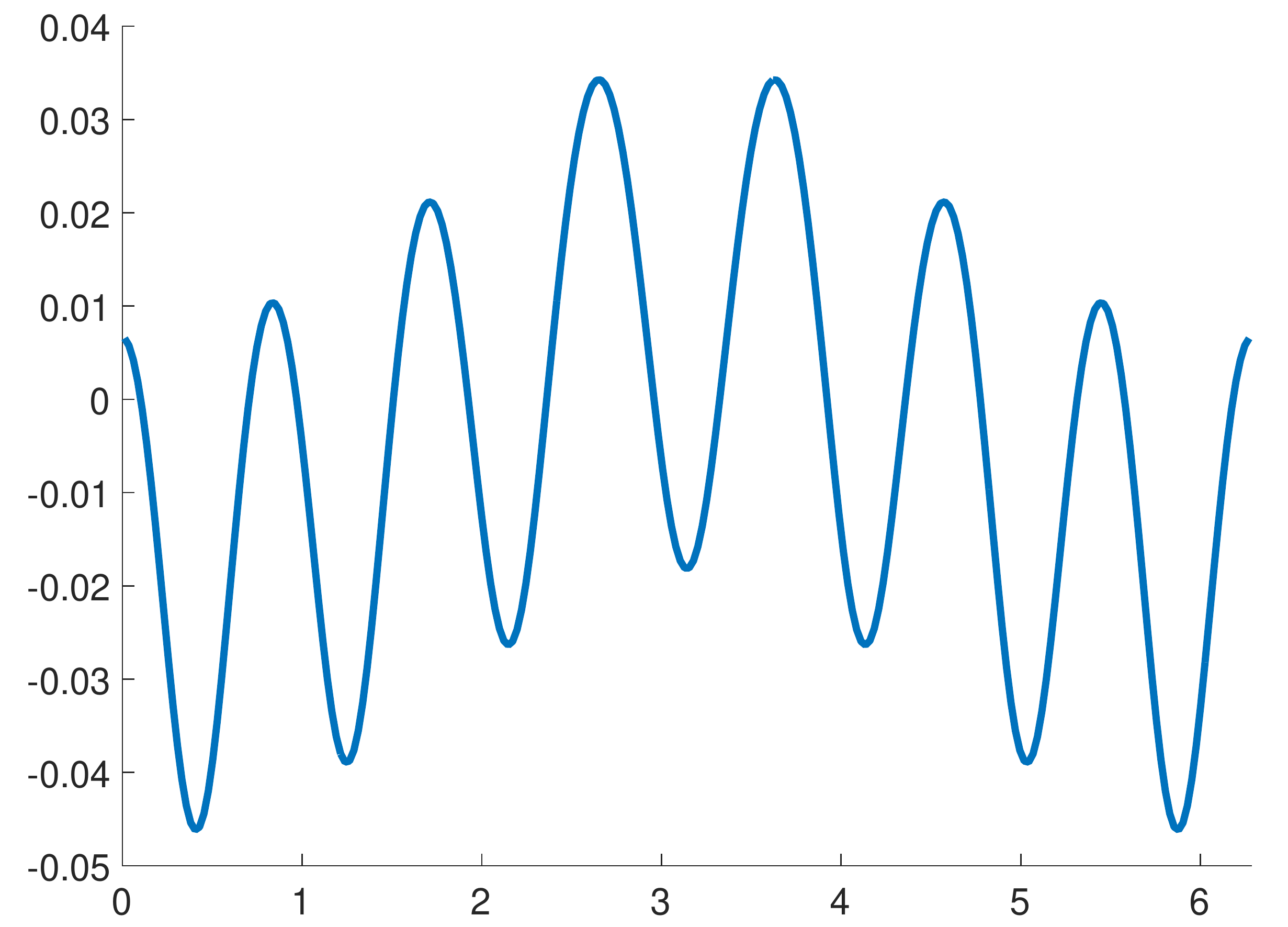}
		\label{fig1and7e}
	}
	\subfloat[]{
		\includegraphics[width=0.39\linewidth]{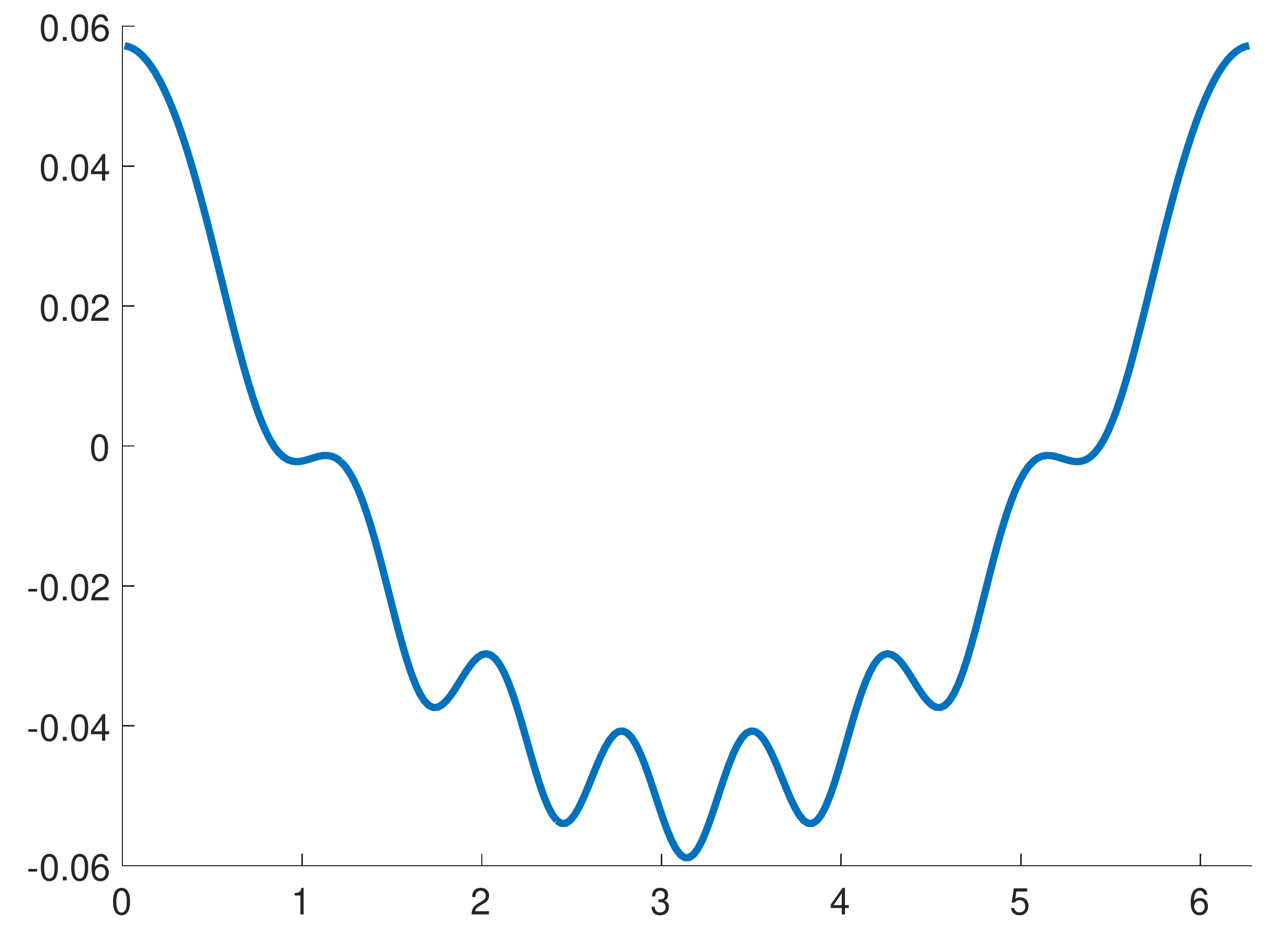}
		\label{fig1and7f}
	}

	\caption{\label{fig1and7}
A Pair of bifurcation branches originating from different but comparable bifurcation points.
The Bond number is given by $T=0.1$. Panel (a) shows the two bifurcation branches,  
and panel (b) shows a close-up of the self-crossing branch.
Panels (c) through (f) show various curves on the self-crossing branch.}
\end{figure}

When $k_1=1$, $k_2=7$, and $T=T(1,7)$, we have seen in the previous section that the two branches bifurcate from the same point. We can therefore view the bifurcation point as a point of connection between these two branches.
Looking at Figure \ref{fig2Dim-1-7} we see that the main branch lies on the right and above the $2\pi/7$-branch; however we know from Formula \eqref{eqBifurcationSpeedCapillary} and Figure \ref{figBifurcationSpeed} that with an increase in $T$, $\mu^*(k_2)$ will increase faster than $\mu^*(k_1)$.
We therefore expect the representations of the two branches in the wavespeed-waveheight plane to \emph{cross} each other at a certain point. 
The question now is: Can we make the two branches \emph{connect}, i.e. can we make small variations in $T$ such that there still exists a point (which was originally at $(0,\mu^*)$) where the two branches share the same wave?
The answer in general is yes, provided a multiplicity condition on the wavenumbers is fulfilled.

\begin{figure}[htb]
	\centering
	
	\subfloat[$T=0.1$]{
		\includegraphics[width=0.39\linewidth]{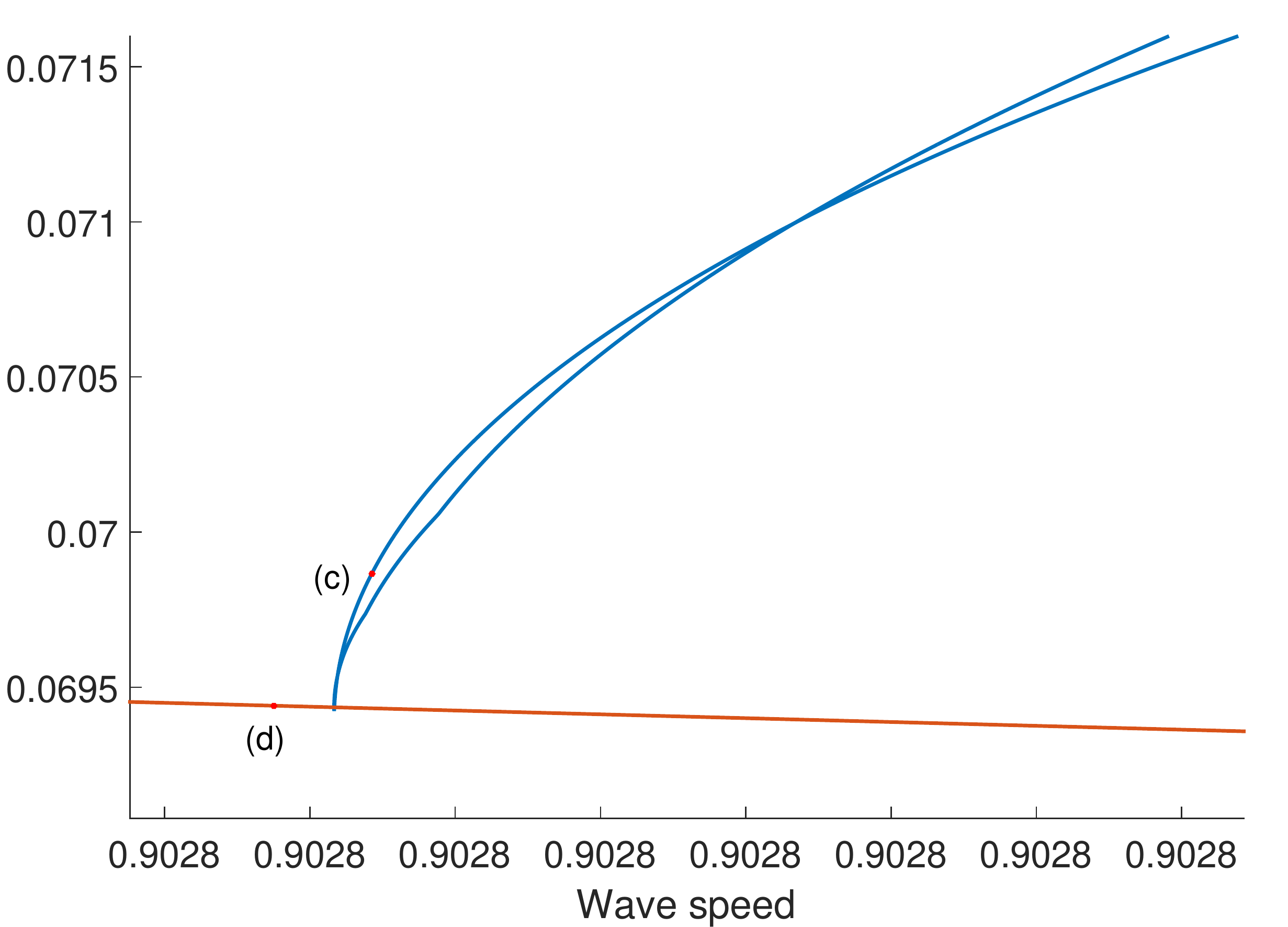}
		\label{fig1and7closeup2}
	}
	\subfloat[]{
		\includegraphics[width=0.39\linewidth]{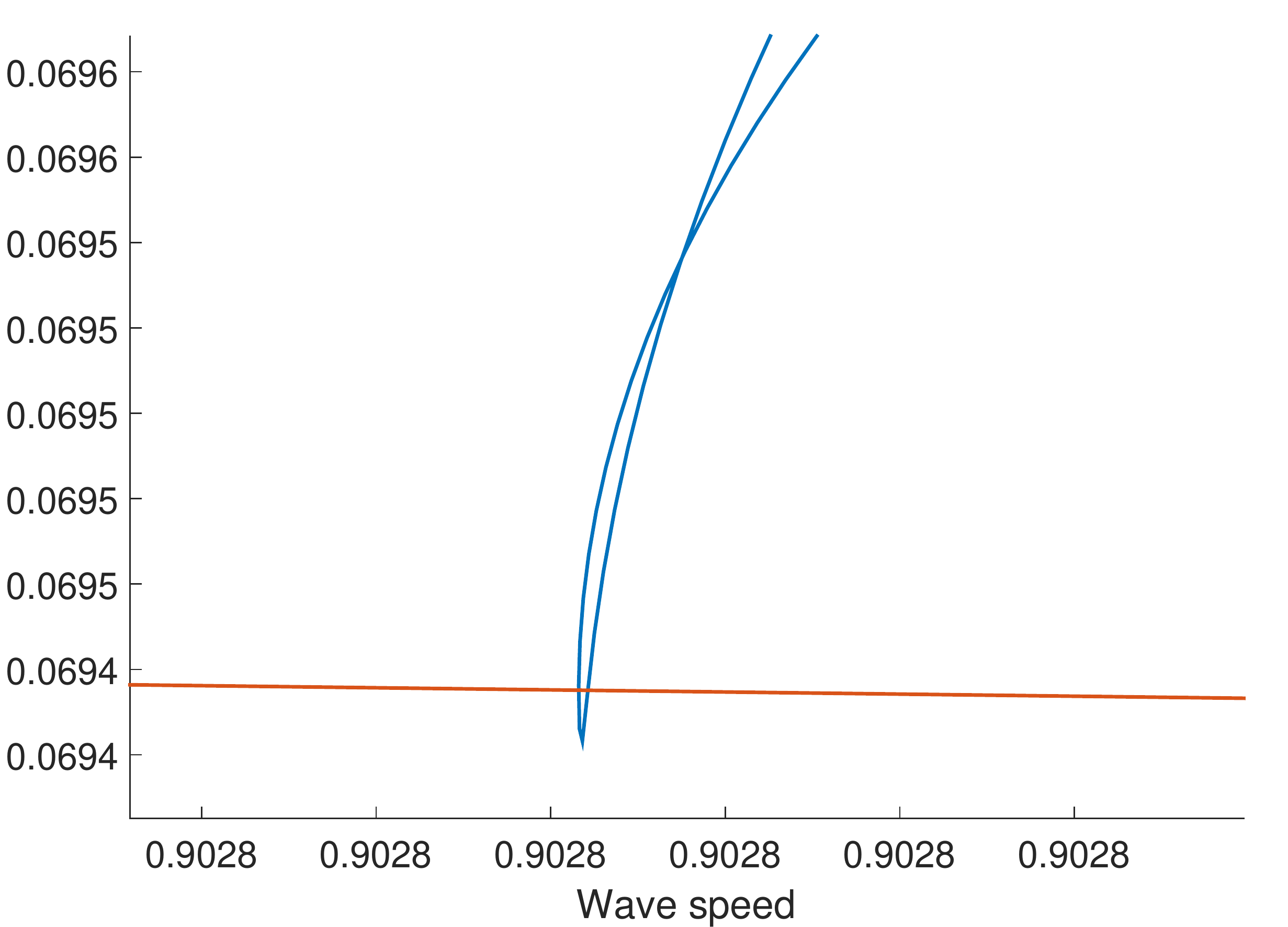}
		\label{fig1and7closeup3}
	}

	\subfloat[]{
		\includegraphics[width=0.39\linewidth]{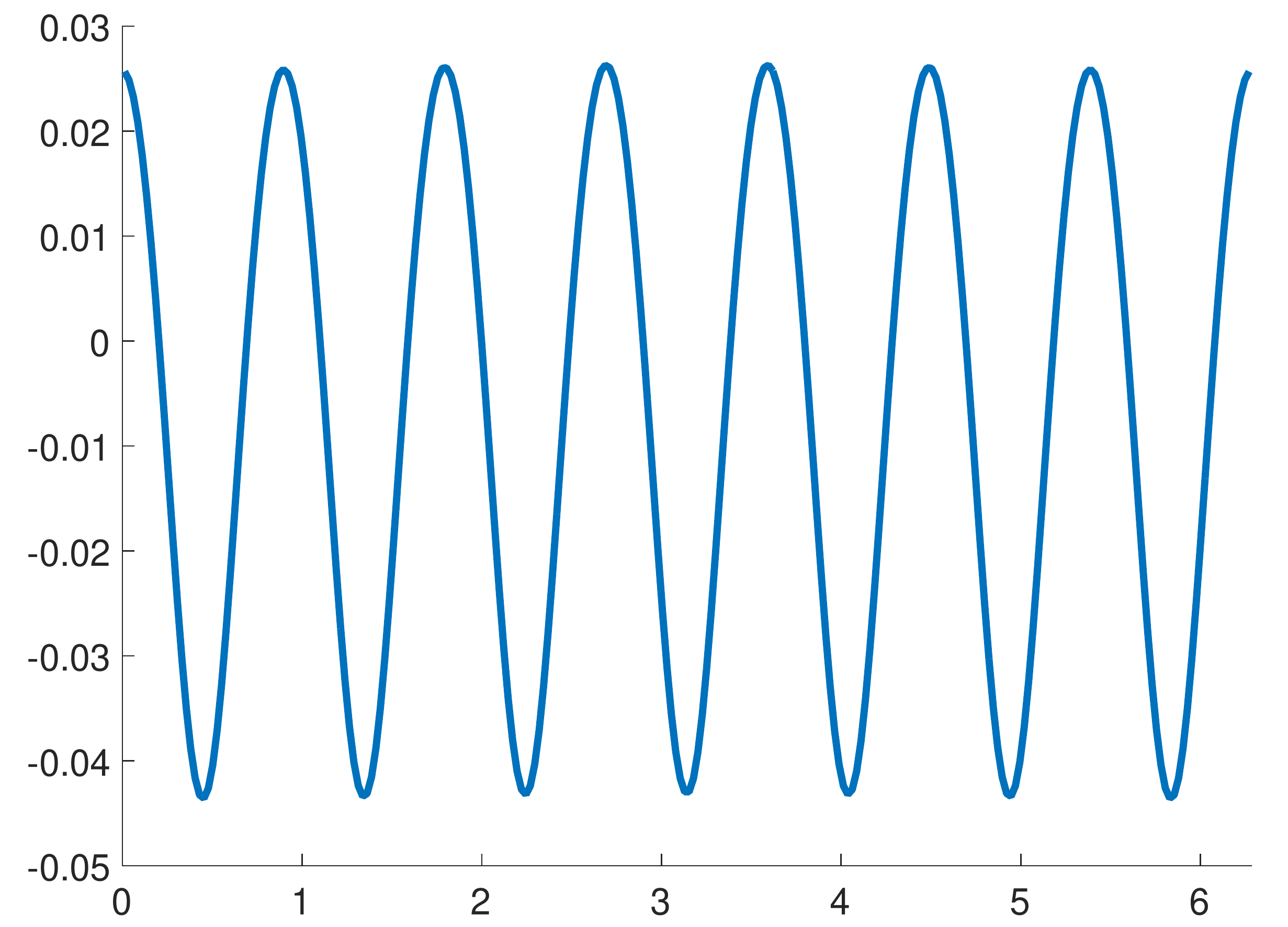}
		\label{fig1and7-connection1}
	}
	\subfloat[]{
		\includegraphics[width=0.39\linewidth]{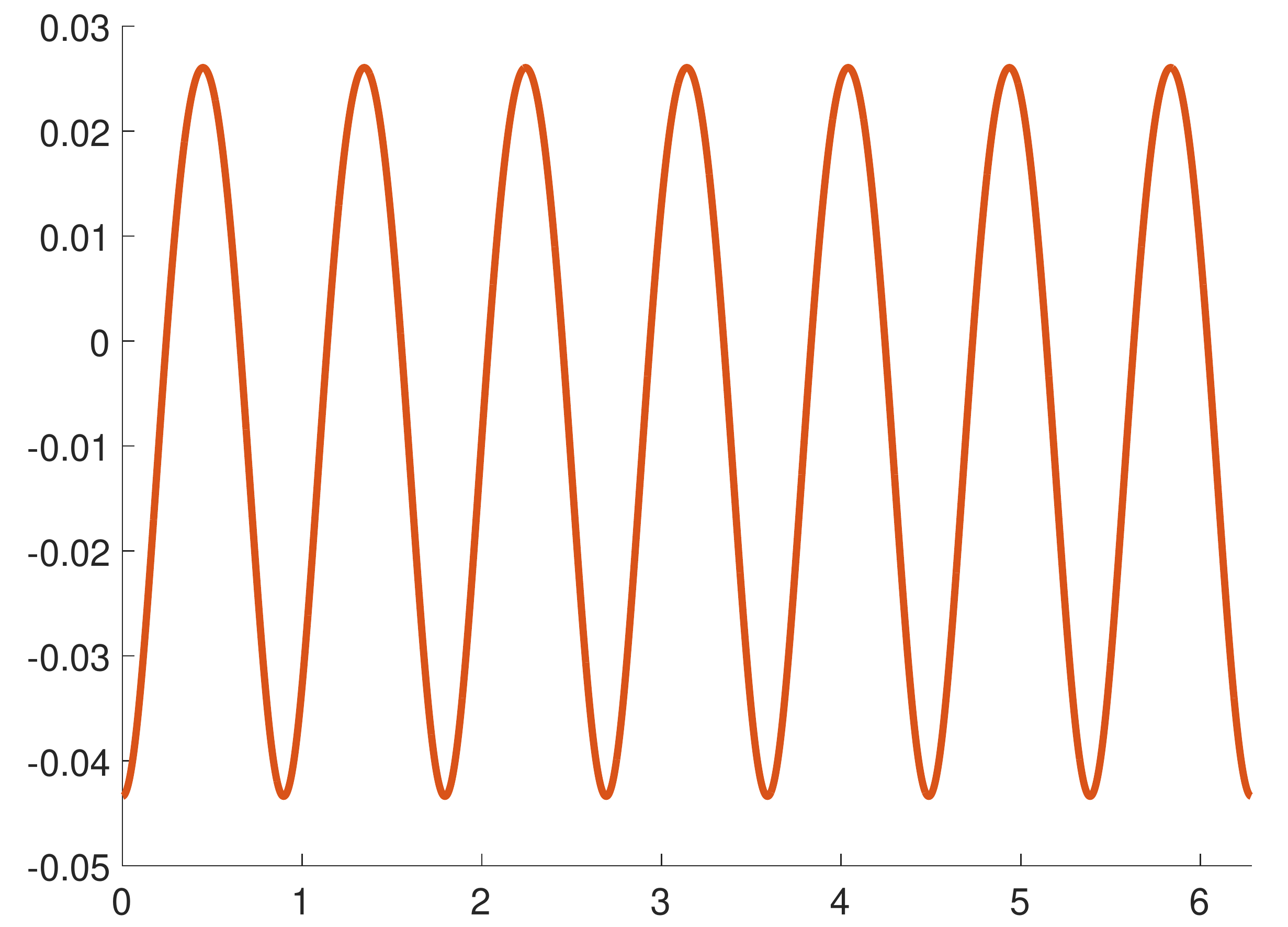}
		\label{fig1and7-connection7}
	}

	\caption{\label{fig1and7Closeup}
Panels (a) and (b) show close-ups of the intersection zone of the two bifurcation branches
shown in Figure \ref{fig1and7}. In panel (b), the secondary bifurcation point where the two branches connect is the left one.
Panel (c) shows a curve on the blue branch, and panel (d) shows a curve on the red branch.}
\end{figure}

Figure \ref{fig1and7} shows the plots for $k_1=1$ and $k_2=7$, with $T=0.1$.
Recall from before that $T(1,7)\approx0.09918$ so now we have $T \approx T(1,7)+0.00082$.
Closeup pictures of the connection point are presented in Figure \ref{fig1and7Closeup}.

As expected, the main branch now starts to the left and below the $k_2$ branch, 
and near the point labelled (c), they cross each other without connecting since they do not share the same solution at that point.
Very similarly to what we have seen in Section \ref{sec2Dim}, the profile of the wave starts as $\cos(x)$ right after the bifurcation point, then loses monotonicity (Figure \ref{fig1and7c}) and rapidly develops seven crests (Figure \ref{fig1and7d}).
The further we go up the branch the more evident is the presence of a ``carrier'' signal like $\cos(x)$ and a high frequency modulation given by the $\cos(7x)$ component: See Figure \ref{fig1and7e}.
The main branch then curves and connects to the $2\pi/7$ one: Figure \ref{fig1and7Closeup} shows two close-ups of the connection point.
While approaching the $2\pi/7$ branch, the main branch crosses itself twice but does not self-intersect.
After that it also crosses the $2\pi/7$ branch, then turns back and actually connects to it; the connection point being the left one in Figure \ref{fig1and7closeup3}.
Near that point we see that the profiles are essentially identical (Figures \ref{fig1and7-connection1} and \ref{fig1and7-connection7}).
After the connection the main branch separates again and moves up, forming a loop (Figure \ref{fig1and7a} and closeup in Figure \ref{fig1and7b}) before continuing in the direction of increasing heights.
Again, there is no self-intersection in the loop, but only a crossing.
After the connection point with the $k_2$ branch, the profiles in the main branch are flipped vertically and the contribution from the $\cos(7x)$ component diminishes until it reaches a situation like the one presented in Figure \ref{fig1and7f}.
The profiles remain essentially unchanged in shape further up in the branch.

\begin{figure}[htb]
	\centering
	\subfloat[$T=T(1,4)+0.0001$]{
		\includegraphics[width=0.39\linewidth]{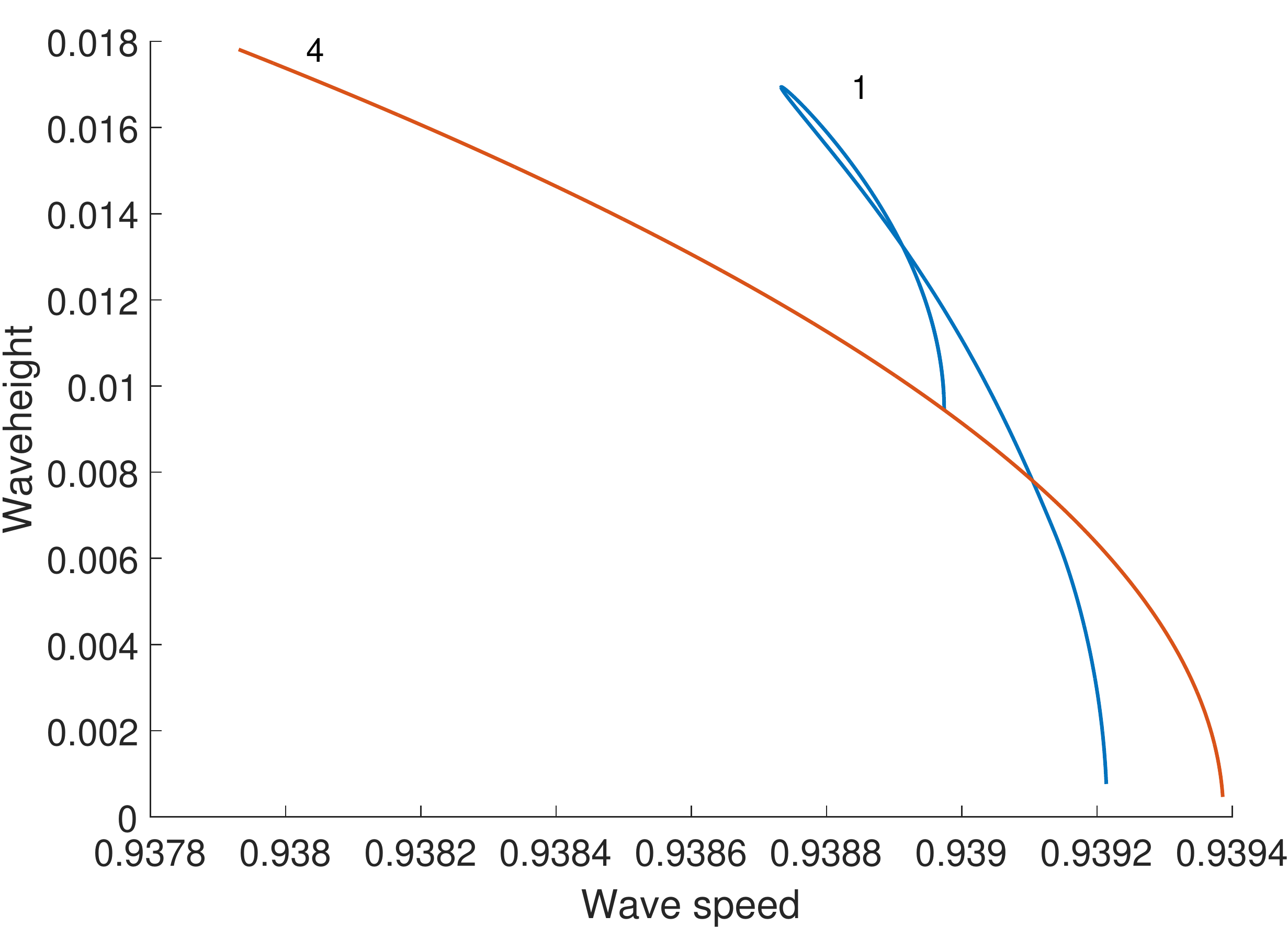}
		\label{figOtherConnectionsa}
	}
	\subfloat[$T=T(1,5)+0.0001$]{
		\includegraphics[width=0.39\linewidth]{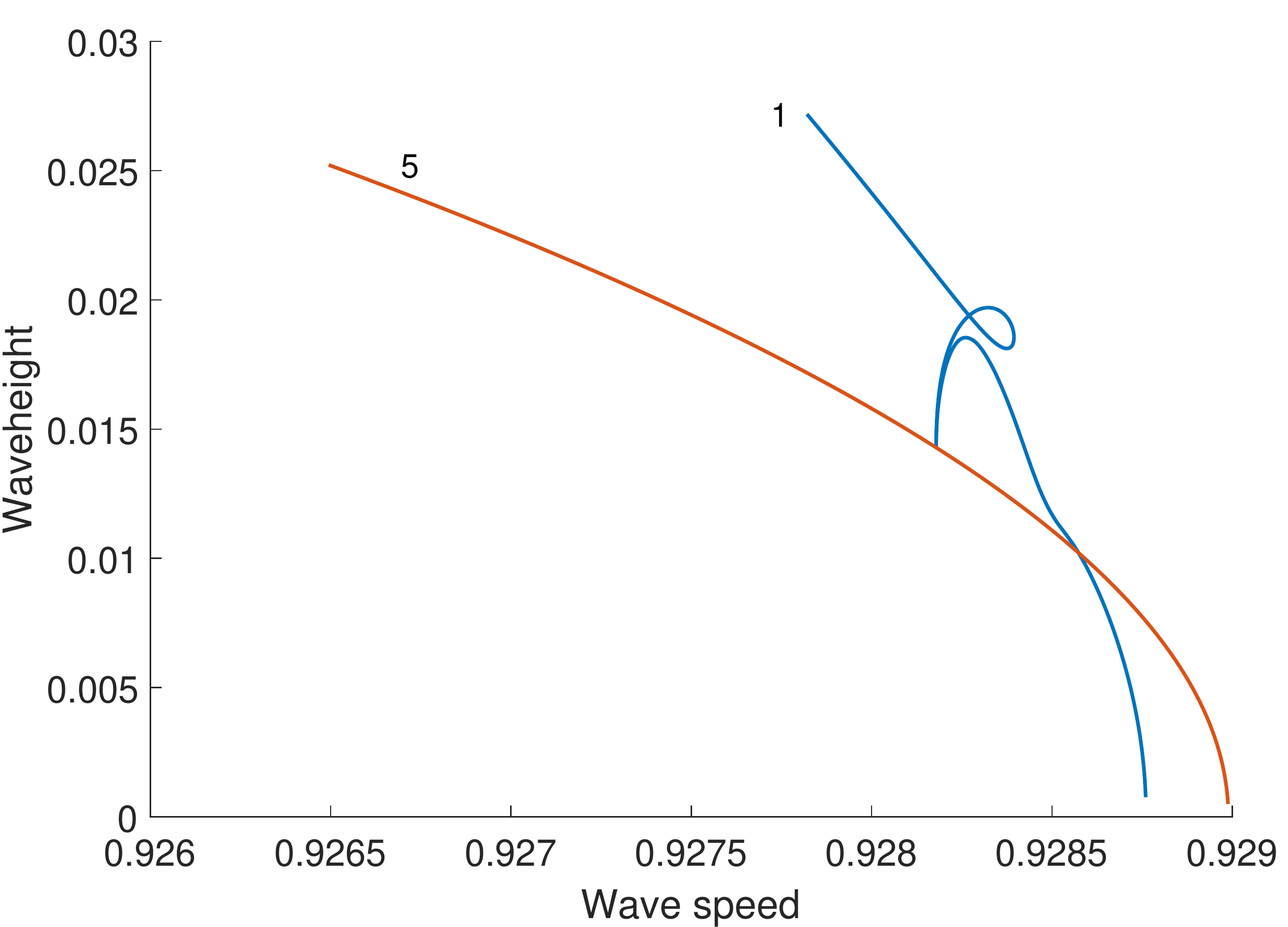}
		\label{figOtherConnectionsb}
	}
	
	\subfloat[$T=T(2,8)+0.0001$]{
		\includegraphics[width=0.39\linewidth]{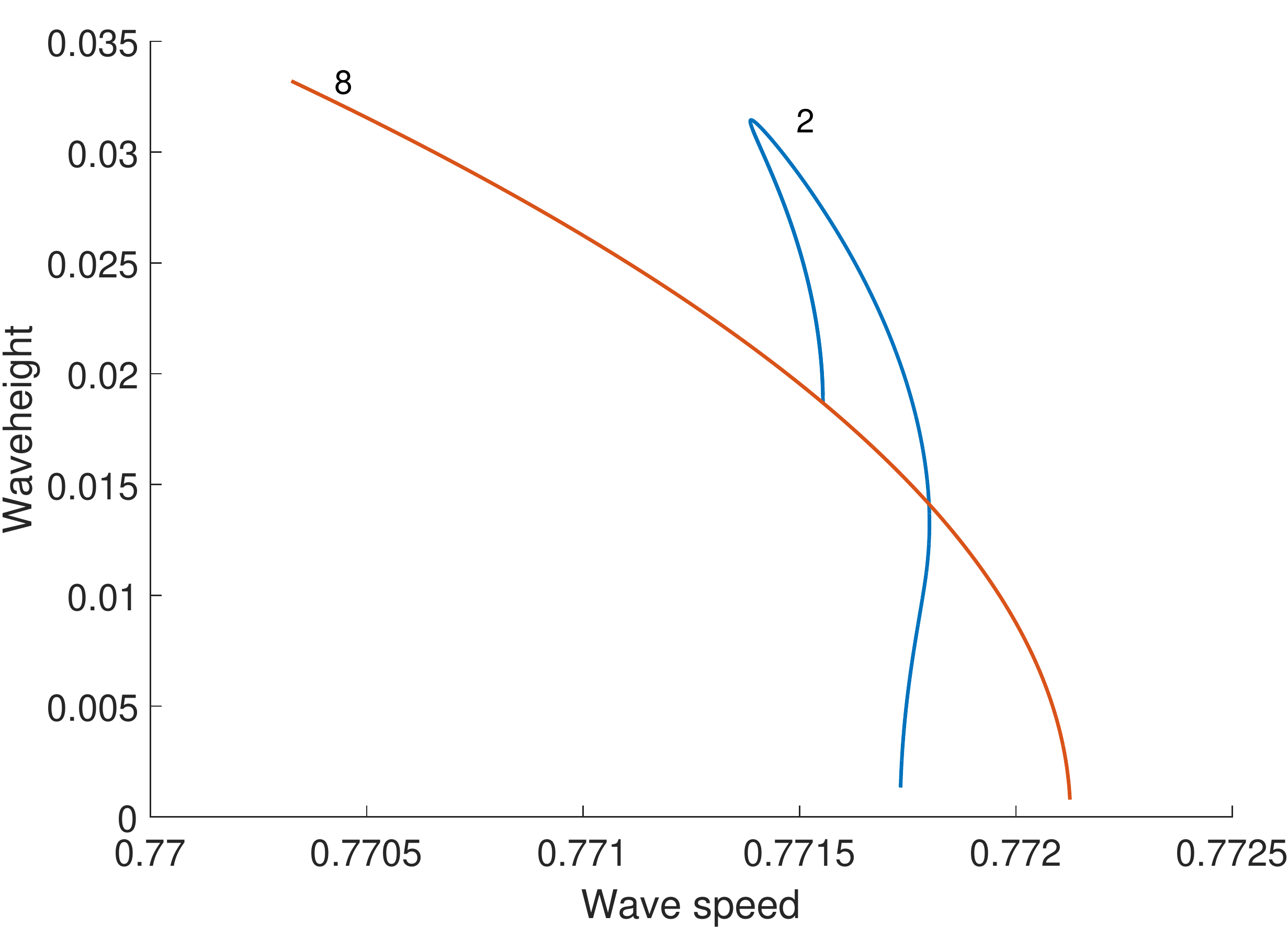}
		\label{figOtherConnectionsc}
	}
	\subfloat[$T=T(2,10)+0.0005$]{
		\includegraphics[width=0.39\linewidth]{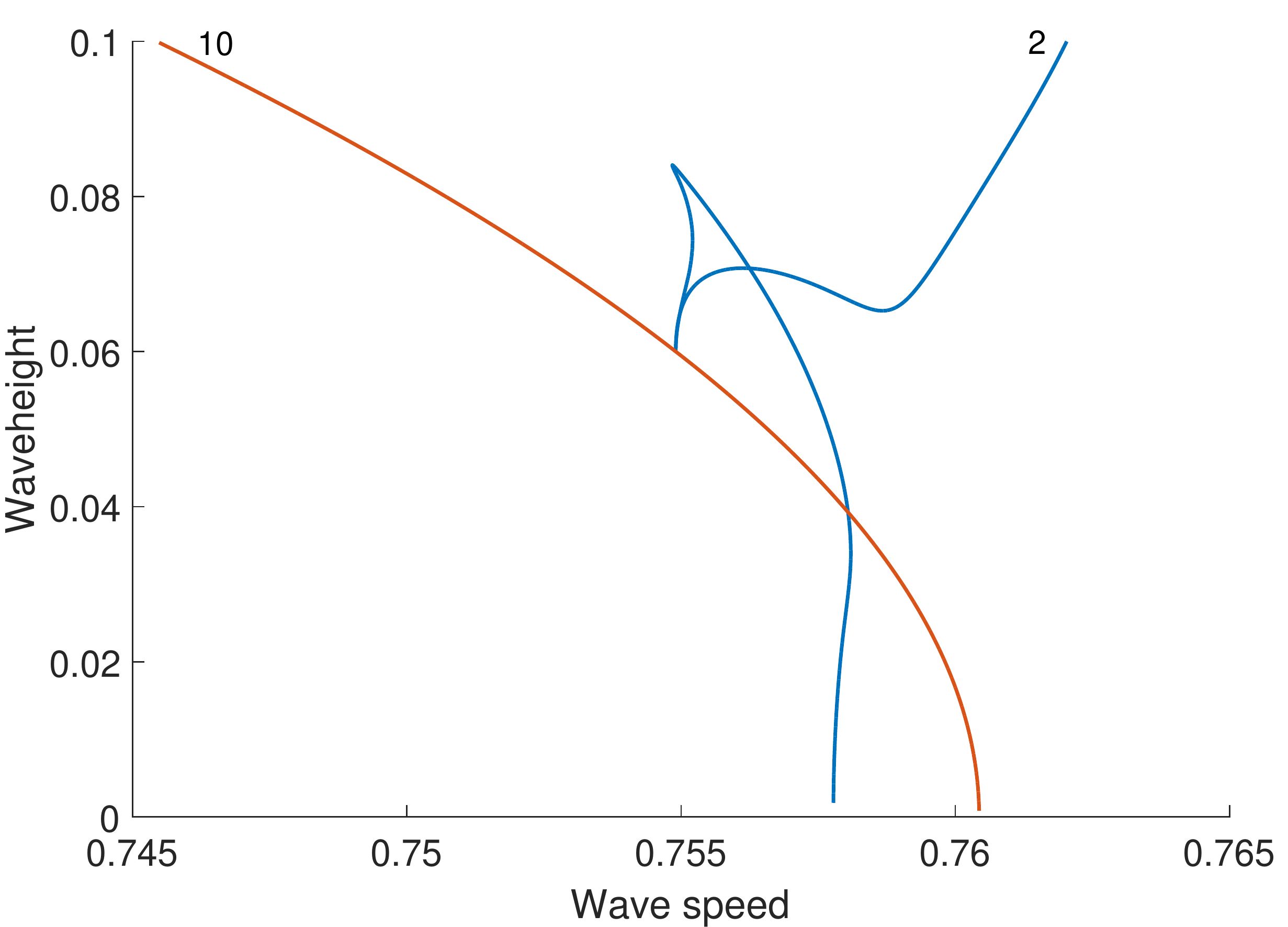}
		\label{figOtherConnectionsd}
	}
	\caption{\label{figOtherConnections}
Various intersecting and self-crossing branches. Panels (b) and (d) feature secondary bifurcations.}
\end{figure}

It is possible to replicate the above picture using any $k_1$ provided $k_2$ is chosen as
\begin{equation}
	\label{eqKCondition}
	k_2 = (4+a)\,k_1,\quad a\in \mathbb{N}_0.
\end{equation}
In particular, our numerical experiments show that if $a$ is odd, and hence $k_2$ is an odd multiple of $k_1$, then the lower-mode branch connects with the higher one, but after the connection it continues and is unbounded as we can see in the previous Figure \ref{fig1and7a}.
If instead $a$ is even, then the $k_1$ branch terminates at the connection point and no further solutions are found: See Figure \ref{figOtherConnections}.

\section{Acknowledgments}
This research was supported in part by the Research Council of Norway through grants 213474/F20 and 231668.
The authors would like to thank Mats Ehrnstr\"{o}m for help in the preparation of this manuscript.
The authors would also like to thank Mathew Johnson and Kyle Claassen for interesting discussions 
on the pseudo-arclength parametrisation.


\end{document}